\documentclass[aps,prb,10pt,twocolumn,showpacs,floatfix,superscriptaddress,amsmath,amssymb]{revtex4-1}

\pdfoutput=1 

\usepackage{url}
\usepackage{graphicx}
\usepackage{dcolumn}
\usepackage{bm}
\usepackage{color}
\usepackage{version}
\usepackage{amsfonts}
\usepackage{amsmath}
\usepackage{multirow}
\usepackage{rotating}
\usepackage{adjustbox}

\newcommand{\ba}{\begin{eqnarray}}
\newcommand{\ea}{\end{eqnarray}}
\def\be{\begin{equation}}
\def\ee{\end{equation}}

\def\vp{\vec{p}}
\def\vP{\vec{P}}

\newcommand{\angstrom}{$\mathrm{\mbox{\AA}} \,$}

\newcommand{\ve}[1]{ \mathbf{#1}}
\newcommand{\vf}[1]{ \vec{#1}}

\excludeversion{details}

\begin{document}

\title{First-Principles Study of the Ferroelectric Properties of SrTaO$_2$N/SrTiO$_3$ Interfaces.}

\author{R. C. \ Bastidas Brice\~no}
\address{Departamento de Ciencias B\'asicas, Facultad de Ingenier\'ia Universidad Nacional de La Plata, La Plata, Argentina.}
\address{Instituto de F\'isica La Plata (IFLP), CONICET, Argentina.}

\author{V. I.\ Fern\'andez}
\address{Departamento de F\'isica, Facultad de Ciencias Exactas, Universidad Nacional de La Plata, Argentina.}
\address{Instituto de F\'isica La Plata (IFLP), CONICET, Argentina.}

\author{E. \ Peltzer y Blanc\'a}
\address{Consejo Nacional de Investigaciones Cient\'ificas y T\'ecnicas (CONICET), Argentina.}

\author{ R. E.\ Alonso}
\address{Departamento de Ciencias B\'asicas, Facultad de Ingenier\'ia Universidad Nacional de La Plata, La Plata, Argentina.}
\address{Instituto de Ingenier\'ia y Agronom\'ia, Universidad Nacional Arturo Jauretche, Argentina.}
\address{Instituto de F\'isica La Plata (IFLP), CONICET, Argentina.}

\begin{abstract}

First-principles calculations based on density-functional theory in the pseudo-potential approach have been performed for the total energy, crystal structure and cell polarization for SrTaO$_2$N/SrTiO$_3$ heterostructures. Different heterojunctions were analyzed in terms of the termination atoms at the interface plane, and periodic or non-periodic stacking in the perpendicular direction. The calculations show that the SrTaO$_2$N layer is compressed along the $ab$-plane, while the SrTiO$_3$ is elongated, thus favoring the formation of P4mm local environment on both sides of the interface, leading to net macroscopic polarization. The analysis of the local polarization as a function of the distance to the interface, for each individual unit cell was found to depend on the presence of a N or an O atom at the interface, and also on the asymmetric and not uniform $c$-axis deformation due to the induced strain in the $ab$-plane. The resulting total polarization in the periodic array was $ \approx 0.54$ C/m$^2$, which makes this type of arrangement suitable for microelectronic applications.

\end{abstract}
\maketitle

\section{Introduction.}

In nature there exist several compounds with chemical formula ABO$_3$ and perovskite structure (for example A=Ca, Sr, Ba or lanthanide, B= Ti, Ta or Nb). For some A and B combinations, compounds with highly desirable physical properties such as superconductivity, ferromagnetism, ferroelectricity, piezoelectricity and giant magneto-resistance have been found\cite{SKIM1999}. These materials have been widely investigated in the last decades due to the scientific and technological interesting properties they present, leading perovskite-type compounds to an extensive use in microelectronics. Some of them present drastic anomalies in their dielectric constant at a given temperature that indicate the occurrence of a phase transition, for example a paraelectric-ferroelectric or else a paraelectric-antiferroelectric phase transition. In the last years AB(O,N)$_3$ oxynitrides have attracted much attention due to their novel electronic functionalities, such as high dielectric constant\cite{YIKIM2004}, visible light absorption\cite{JANSEN2000},  photocatalytic activity and colossal magneto-resistance\cite{KASAHARA2002, MYANG2010}. They are formed by the partial substitution of oxygen by nitrogen atoms into the anionic network of the ABO$_3$ perovskites.  The consequent increase of negative charges is compensated with cationic substitutions in the B site (for example Ti by Ta) \cite{MARCHANDPAT, MARCHAND1986, PORS1988, MARCHAND1991}. The replacement of O by N may change the color of the initially white powders, so, the oxynitrides have potential use as safe colored pigment material able to replace currently used contaminant heavy metals (Cr, Cd, Pb, etc.)\cite{JANSEN2000}. The change in the resulting optical properties has fueled a number of recent studies. However, the changes in bonding and structure that come along with the introduction of nitride ions may also give rise to interesting dielectric behavior. The aliovalent substitution provides a mechanism for enhancing the dielectric polarizability through the substitution of more polarizable ions into the lattice\cite{SHANNON1993}. Furthermore, the reduced electronegativity of the nitride ion, with respect to the oxide ion, will tend to increase the covalency of the cation-anion bonds\cite{YIKIM2004}. The increased covalency of the bonding can in turn increase the likelihood of cation displacements through a second-order Jahn-Teller distortion of the $d^0$ cation\cite{HALASYAMANI1998, BERSUKER2001, BURDETT1995}. In literature, such displacements are commonly associated with the origin of ferroelectricity in many cases. On the other hand, the mixed occupancy of the anion sites in oxynitrides, AM(O$_{1-x}$N$_x$)$_3$, provides a condition like that found in relaxors, as the polarized octahedral cations will experience random chemical environments in the absence of complete O/N ordering. It is, therefore, an interesting issue to examine whether the oxynitride perovskites possess intrinsically high $\kappa$ and relaxor-like properties.

The electronic functionalities of perovskite oxynitrides may be influenced by the geometrical configuration of the O and N atoms around their metal cations. For ABO$_2$N-type oxynitrides, in which each B cation is surrounded by four O and two N ions forming a BO$_4$N$_2$ octahedron, there are two possible anion configurations: the two nitrogen ions can occupy either adjacent (cis-type) or opposite (trans-type) sites. Researchers have argued that the dielectric properties of ABO$_2$N are related to anion arrangement: for example, Page et al. suggested that ferroelectricity in trans-type anion-ordered ATaO$_2$N (A=Sr, Ba) phases may be caused by the off-center displacement of Ta ions\cite{PAGE2007}. They investigated this concept by theoretically studying the stability of phases with different nitrogen arrangements and space groups. However, trans-type phases in this system are less energetically stable than cis-type phases, and bulk SrTaO$_2$N (STN) specimens have been confirmed to exhibit cis-type configurations in an I4mcm centrosymmetric space group\cite{MYANG2011, ZHANG2011}. However, different ferroelectric and relaxor regions were recently detected in thin films samples of STN epitaxially grown on a SrTiO$_3$ (STO) substrate\cite{OKA2014, ZHU2014}. The lattice mismatch between the compound and the substrate produces strain on the oxynitride in the plane parallel to the interface thereby reducing the in-plane lattice constant when it is grown as thin film. This induced strain favors the stabilization of a trans-type non centrosymmetric polar P4mm ferroelectric structure\cite{PAGE2007, OKA2014}. Also, J. H. Haeni et al. suggested that strain and substrate clamping induces in epitaxial STO films structures that are not present on the bulk material. For the case of tensile strain, a phase transition from a the tetragonal P4mm high temperature phase to the orthorhombic Cmcm low temperature phase may occur\cite{HAENI2004}.

Computational quantum simulations have proven to be a suitable tool for understanding the microscopic processes that drive the physical properties of materials. Beyond the simplifications that are made to model the systems, due to the complexity of real materials (especially in disordered and ceramic materials), they can lead to a better understanding of the individual processes involved. Some previous studies on phase stability in bulk-strained STN and STO have been published\cite{PAGE2007, OKA2014, HAENI2004, ALONSO2018, HINUMA2012}. Nevertheless, the behavior of the SrTiO$_3$/SrTaO$_2$N interface opens new questions regarding the mutually induced strain and cell deformations, together with the possible coupling of the dipole moments above and below the interface. The bulk lattice mismatch ($a_{\mbox{{\tiny{STO}}}}$= 3.905 \angstrom, $a_{\mbox{{\tiny{STN}}}}$= 4.0271 \angstrom) produces a tensile strain for the STO layer and a compressive strain on the STN layer when the interface is formed. Accordingly, both compounds could stabilize in the ferroelectric P4mm phase. In this work, we study different configurations of SrTiO$_3$/SrTaO$_2$N interfaces within the Density Functional Theory (DFT)\cite{KOHN1965}. We will analyze the ferroelectric distortion in the tetragonal induced P4mm phases above and below the interface, due to compressive strain on the side of SrTaO$_2$N and tensile strain corresponding to the SrTiO$_3$, in order to compute the resulting net dipole moment. The paper is organized as follows: in Section \ref{sec:configurations} the configurations of the different constructed interlayers are explained. In section \ref{sec:method} the method used for calculations is described. In Section\ref{sec:bulk} the comparative results of the calculations of the pure STN and STO vs. the SC’s systems are shown. In Section \ref{sec:polarization}  a detailed analysis of the cell and sub-cell polarization is presented. In Section \ref{sec:conclusions}  the summary and conclusions. Finally, we include Supplementary Information with a detailed information of the atomic positions and polarization contributions per atom for the SCs in study.

\section{The system under study.}
\label{sec:configurations}

\begin{figure*}
\begin{center}
\begin{adjustbox}{width=1\textwidth}
\includegraphics{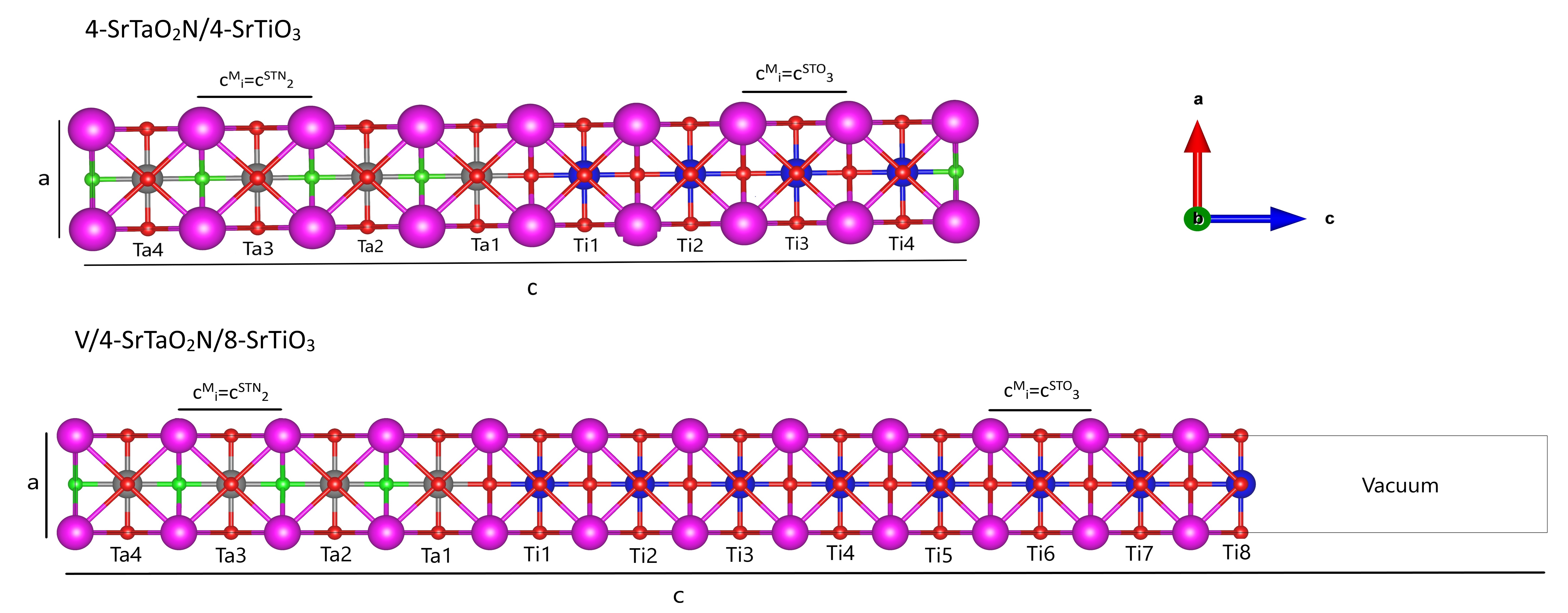}
\end{adjustbox}
\end{center}
\caption{Color online. Schematic atomic distribution in 4-SrTaO$_2$N/4-SrTiO$_3$ (above) and V/4-SrTaO$_2$N/8-SrTiO$_3$ (below) SCs.}
  \label{fig:structures}
\end{figure*}

\begin{figure}[h!]
\begin{center}
\includegraphics[width =0.5\textwidth]{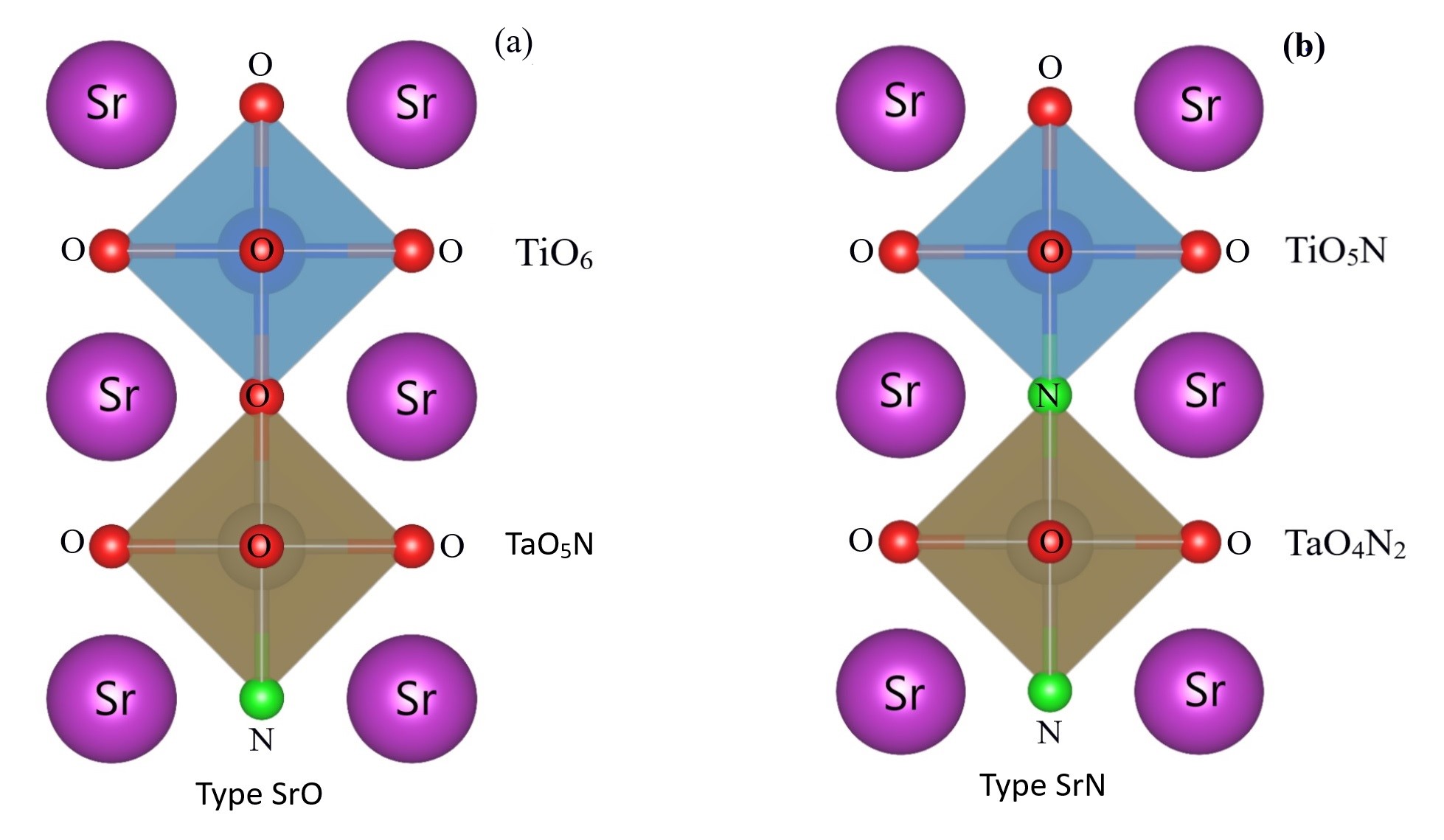}
\end{center}
\caption{Color online. Detailed view of the different termination modes at the interface (central horizontal Sr's plane). SrO-type termination corresponds to TaO$_5$N/TiO$_6$ octahedra (a), while SrN-type termination corresponds to TaO$_4$N$_2$/TiO$_5$N octahedra (b).}
  \label{fig:octahedra}
\end{figure}

Bulk STN have been reported to exhibit cis-type configuration with I4mcm space group at room temperature (RT),with lattice constants $a$= 5.70251(6) \angstrom  and $c$= 8.05420(16) \angstrom  (using the pseudo-cubic parameters $a \approx$ 4.032 \angstrom  and $c \approx$ 4.0271 \angstrom \cite{YIKIM2004}). On the other hand, at room temperature STO crystallizes in the simple cubic perovskite structure with space group Pm3m and lattice parameter 3.905 \angstrom. For the case of an interface consisting of a STN thin film over a STO substrate, a substantial lattice mismatch (3.28\%) exists during hetero-epitaxial deposition due to the difference in the in-plane lattice constants. As a result, the substrate becomes subjected to high tensile strain, which in turn may cause an alteration of the lattice symmetry of STO from cubic to tetragonal, and this results in the consequent appearance of Raman-inactive first-order phonons\cite{ZHU2014}. Also, at the STN side, a compressive epitaxial stress may stabilize a ferroelectric trans-type P4mm phase, as suggested by Page et al.\cite{PAGE2007}.	In order to study the electrical polarization induced by the tensile (compression) stress in STO (STN) as a function of the distance of the respective sub-cell unit (constituted by each individual perovskite cell structure) to the interface plane, supercells (SC) were made with different sizes and configurations. For this purpose, n-SrTaO$_2$N/n-SrTiO$_3$ (n = 3, 4 and 5) SCs were constructed, stacking n sub-cells of STN over n sub-cells of STO in the 001 direction. An example of these structures for n=4 can be observed in Figure \ref{fig:structures} (above). The periodic boundary condition transforms this geometry in a periodic multilayer .../STN/STO/STN/... with planar interfaces. If the interface is constructed along the $ab$-plane at the Sr atoms level, then, in the center of the four Sr square plane there will be either an O or a N atom. If there is an O atom (SrO interface-type), the STO sub-cell just above the interface has a normal TiO$_6$ octahedron, whereas the STN sub-cell immediately below the interface has a TaO$_5$N octahedron (See Fig. \ref{fig:octahedra} (a)). In order to maintain the stoichiometry, in the next interface, generated by the periodic boundary conditions, there will be a N atom in the middle of the Sr square (SrN interface-type) (See Fig. \ref{fig:octahedra} (b)) Thus, the STO sub-cell immediately below the interface has a TiO$_5$N octahedron and the STN sub-cell just above has a TaO$_4$N$_2$ octahedron, being this last the configuration for trans-type oxynitrides. Therefore, this periodic multilayer array allows the analysis of the electronic behavior of the two types of interface-termination cases (SrO and SrN) simultaneously.
Also, in order to simulate a unique interface between a STO substrate and a STN thin film to model the experimental study of Ref. \cite{OKA2014}, the V/m-SrTaO$_2$N/n-SrTiO$_3$ cells were also built, with mxn = 4x8 and 5x10 (Figure \ref{fig:structures} (below)). These SCs have twice the unit formulas of STO with respect to those of STN, and a V vacuum separator layer 10 \angstrom  thick in the [001] direction. The vacuum layer prevents the induction of dipole moments and strain in the STO layer from below by the STN layer due to periodicity, as it occurs in the previous multilayer scheme. In addition, the greater number of STO sub-cells compared with those of STN, better reflects the experimental situation mentioned\cite{OKA2014}. The V/m-SrTaO$_2$N/n-SrTiO$_3$ SCs were built with two types of possible terminations at the interface, that is, with an O or a N atom at the separation plane of the interface. They correspond to SrO or SrN-type respectively (As seen in Fig. \ref{fig:octahedra}).

For all the SCs in this work, an initial tetragonal configuration was used for both layers, where the initial lattice constant a has been chosen as an intermediate value of the corresponding to bulk STO and STN. For the STO layer, the initial positions were selected as a centrosymmetric configuration (4/mmm), whereas for the STN layer, a non-centrosymmetric (4/mm) configuration was selected, being the Ta atoms slightly displaced outside of the symmetry plane. All the initial sub-cells of each layer were built with the same configuration. From these initial configurations, the whole atomic positions and cell constants were relaxed, as described below.
	
Tests were performed with a different arrangement of the N atoms in the STN shell: trans-type N atoms parallel to the $ab$-plane of the interface resulted in orthorhombic final structures, and with final total energies higher than those corresponding to the structures depicted above. For example, for the 3-SrTaO$_2$N/3-SrTiO$_3$ SC, the energy difference between the trans-type N atoms parallel and perpendicular to the $ab$-plane resulted in 0.034 eV, which is greater than the precision of the calculation method. Then, only the $c$-ordered trans-type N atoms was considered.

\section{Method of calculation.}
\label{sec:method}

{\it{Ab-initio}} electronic structure calculations were used to determine the self-consistent potential and the charge density of the different configurations analyzed, and from these first-principles calculations, the cell polarization was obtained. The calculations have been performed using the QUANTUM ESPRESSO code\cite{QUANTUMESPRESSO}, which is based on the density functional theory (DFT)\cite{KOHN1965}.

For the electronic exchange-correlation potential the generalized gradient approximation (GGA) based on the Perdew-Burke-Ernzerhof expression was used\cite{PERDEW1996}. The electron-ion interaction was treated by using Vanderbilt ultrasoft pseudopotentials\cite{VANDERBILT1990}, with the following valence electronic configuration, Sr($4s^2 \,\, 5s^2 \,\, 4p^6 \,\,5p^0$ ), Ti($3s^2 \,\, 4s^2 \,\, 3p^6 \,\, 4p^0 \,\, 3d^2$), Ta($5s^2 \,\, 6s^2 \,\, 5p^6 \,\, 6p^0 \,\, 5d^3$), N($2s^2 \,\, 2p^3$) and O($2s^2 \,\, 2p^4$). Wave functions were expanded by plane waves with a kinetic energy cutoff of 75 Ry and an energy cutoff of 600 Ry for the charge density. The irreducible Brillouin zone was sampled using the Monkhorst-Pack scheme with a 6x6x1 mesh\cite{MONKHORST1976}. In order to calculate the density of states (DOS) the tetrahedron method has been used with a denser 20x20x5 mesh\cite{LEHMANN1972}, whereas the atomic-projected DOS were calculated by the Lowdin populations\cite{SANCHEZ1995}. For the analysis of the ionic and electronic contribution to the Polarization, the Modern Theory of Polarization formalism according to R.D. King-Smith and D. Vanderbilt\cite{KING1993}, and the Born's Effective Charge Tensor Method were used\cite{NEATON2005, ROY2010}.

\section{Structural and electronic properties.}
\label{sec:bulk}

In order to check the reliability of this theoretical approach and the applicability of the selected pseudopotentials, it was first applied to determine the well-established properties of pure bulk STN and STO. The obtained results were compared with calculations performed using the all-electron full-potential linear augmented plane wave plus local orbital (FP$\_$ APW) method in the scalar relativistic version\cite{SJOSTEDT2000, MADSEN2001, COTTENIER2002}, with the WIEN2k implementation\cite{WIEN2K}, using the same exchange-correlation potential. Thus, before proceeding with the study of the different SCs, a summary of the main results obtained for the pure systems is given below.

{
\begin{table*}
\begin{center}
\begin{adjustbox}{width=1\textwidth}
\begin{tabular}{|c|c|c|c|c|c|c|c|c|c|c|}
\hline
 {\bf{System}} &{\bf{ Method}} & ${a}\, \, (\mathrm{\mbox{{\tiny{\AA}}}} )$ & ${c}\, \,(\mathrm{\mbox{{\tiny{\AA}}}} )$ & $\mbox{{\bf{Volume}}}\atop{(\mathrm{\mbox{{\tiny{\AA}}}^3} )}$
  & O $x$ & $\mbox{{\bf{Ta-N}}}\atop{(\mathrm{\mbox{{\tiny{\AA}}}} )}$ & $\mbox{{\bf{Ta-O}}}\atop{(\mathrm{\mbox{{\tiny{\AA}}}} )}$ & $\mbox{{\bf{Sr-O}}}\atop{(\mathrm{\mbox{{\tiny{\AA}}}} )}$ & $\mbox{{\bf{Sr-N}}}\atop{(\mathrm{\mbox{{\tiny{\AA}}}} )}$
    & ${\mathbf{\Delta}}${\bf{E}} (eV)
  \\ \hline
\multirow{4}{4em}{STN I4/mcm}        & WIEN2k  & 5.586        &  8.2152    & 256.34 & 0.801 & 2.054 & 2.016 & 2.584 & 2.793 &    \\ \cline{2-11}
        & QE  & 5.643 & 8.218 & 261.66  &  0.796    & 2.055 & 2.028 & 2.621 & 2.821 &    \\  \cline{2-11}
      & $\mbox{Experimental}\atop{\mbox{Ref. \cite{YIKIM2004}}}$   & 5.70251(6) & 8.05420(16) & 261.912(3) &   &   2.01355(4)  & 2.019(5) & &  &\\ \cline{2-11}
 & $\mbox{Experimental}\atop{\mbox{Ref. \cite{GUNTHER2000}}}$   & 5.7049(3) & 8.0499(5) & 261.99(4) &  0.7721(4)  &   2.012  & 2.025 & 2.726 & 2.852 & \\ \hline
\multirow{2}{4em}{STOc}  & QE  & 3.939&  & 61.12  &    &  &  &  &  &  0 \\ \cline{2-11}
      & $\mbox{Experimental}\atop{\mbox{Ref. \cite{HAENI2004}}}$   & 3.905 &  & 59.547 &   &     &  & &  &\\ \hline
   $\mbox{STOtnc}\atop{\mbox{P4mm}}$  & QE  & 3.9285 & 3.9865 & 61.52  &    &  &  &  &  &  -0.0006 \\  \hline
     $\mbox{STOtc}\atop{\mbox{P4mm}}$& QE  & 3.939 & 3.938 & 61.10  &    &  &  &  &  &  -0.0007 \\  \hline
\end{tabular}
\end{adjustbox}
\caption{\label{tab:parameters}
  Experimental and calculated lattice constants, volume, O $x$ atomic positions and interatomic distances after full relaxation of the bulk STN and STO structures. The total energy for STO is referred to that of the STOc structure.
}
\end{center}
\end{table*}
}

{\bf{- I4/mcm STN.}}

Input structures for I4/mcm-STN (space group 140) were constructed according to the experimental data from Ref. 22.  In this structure, the atomic Wyckoff positions are Ta $4c$, Sr $4b$, N $4a$ and O $8h$. This last one has fractional coordinates ($x$, 1/2+$x$, 0). The only degrees of freedom in the structure are $a$, $c$ and $x$.

For the WIEN2k calculations, the energy cut-off criterion was R$_{\mbox{\tiny{mt}}}$K$_{\mbox{\tiny{max}}} = 8$ for the system (R$_{\mbox{\tiny{mt}}}$ denotes the smallest muffin-tin radius and K$_{\mbox{\tiny{max}}}$ the largest wave number of the basis set). Integration in reciprocal space was performed using the tetrahedron method, considering 300 {\it{k}}-points in the full Brillouin zone (BZ), which are reduced to 10 {\it{k}}-points in the irreducible wedge of the BZ (IBZ). In Table \ref{tab:parameters}, $a$, $c$, and $x$ parameters obtained after relaxation are presented. Both all-electron and pseudopotential approaches show a good agreement with the experimental data, although the cell volume, the oxygen x position, the Ta-O, Sr-O and Sr-N distances are closer to experimental data for QE calculations. We take this as a very good validation of the method and the pseudopotential selection.

{\bf{- Pm3m STO.}}

For bulk STO, calculations were performed in the cubic Pm3m phase (STOc) and in the tetragonal, both centrosymmetric (I4mcm) and non-centrosymmetric phases (P4mm) (STOtc y STOtnc respectively). The obtained results are shown in Table \ref{tab:parameters}. For the cubic phase (STOc) the lattice constant after optimization resulted in a cell volume 2.6\% higher than the experimental value. This fact lies within the normal overestimation of volume calculations using the GGA approximation. In the optimization of the lattice constants and atomic positions for the STOtc structure, the final a and c values resulted almost the same, resulting in a quasi-cubic cell, in very good agreement with the experimental value of $a$. Whereas, for the STOtnc, it was obtained a tetragonal distortion with $c/a$= 1.015 with the consequent displacements of cations and anions in opposite direction, that give rise to an electrical polarization. The total energy differences between the three structures are around $7 \times 10^{-4}$ eV, which is not significant because such differences are within the calculation method error and moreover, it is well known that quantum fluctuations play an important role in SrTiO$_3$ preventing this compound to make a phase transition to a ferroelectric structure.

\begin{figure}[h!]
\begin{center}
\includegraphics[width= 0.5\textwidth]{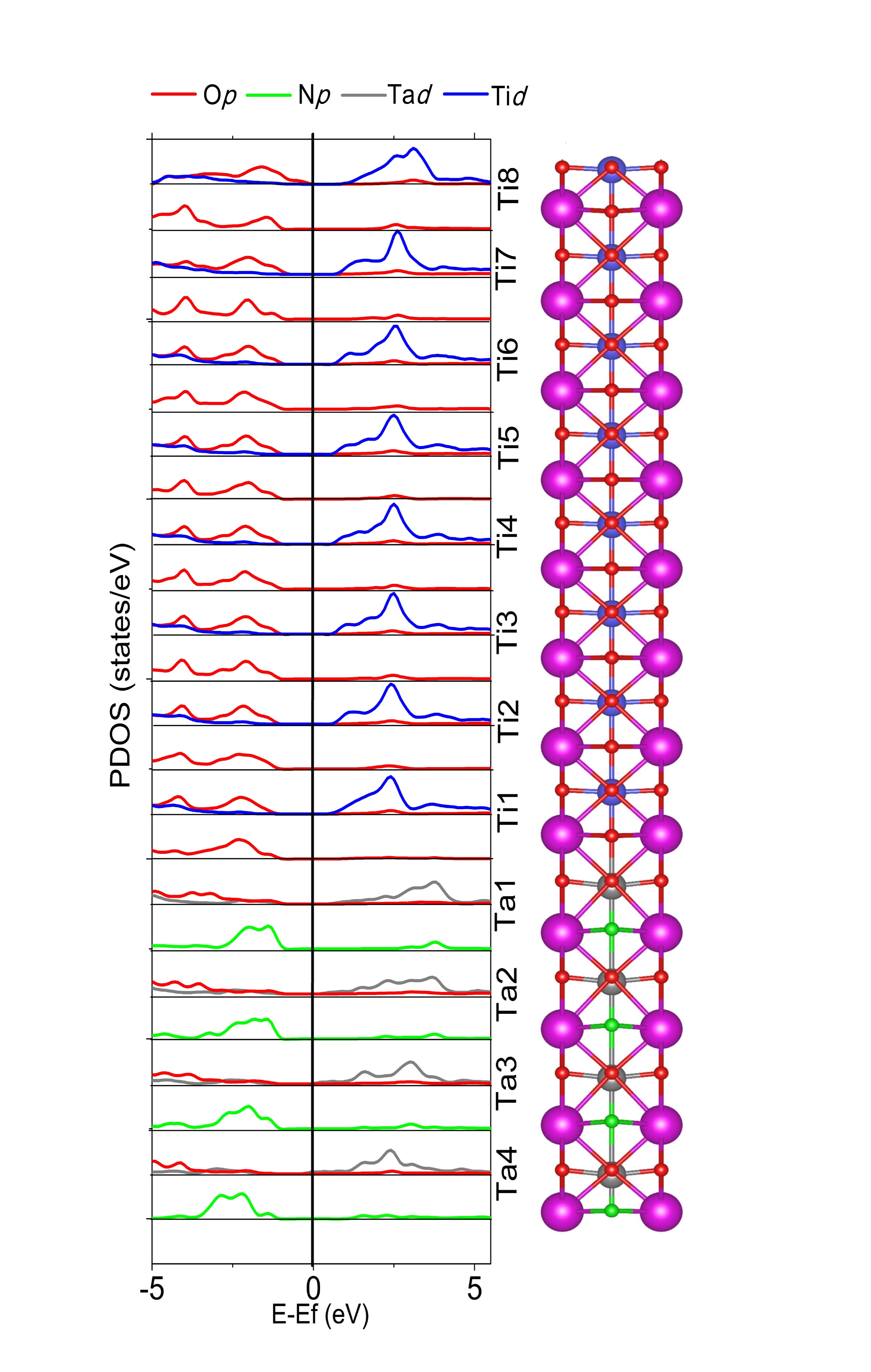}
\end{center}
\caption{Partial density of states for Ta-$d$, Ti-$d$, O-$p$ and N-$p$ atoms calculated for the V-4x8 SC. The corresponding atoms are aligned with the graph. The energy scale is referred to the Fermi level.}
\label{fig:PDOS}
\end{figure}

Once the validation of the method was carried out on the pure compounds, the relaxations of the lattice constants and atomic coordinates of the SCs corresponding to the different n-SrTaO$_2$N/n-SrTiO$_3$ and V/m-SrTaO$_2$N/n-SrTiO$_3$ were performed. To analyze the internal behavior of the individual sub-cell units of the SrBO$_x$N$_y$ perovskites (B = Ti or Ta,  $x + y$ = 3), the lattice constants of the individual sub-cell were considered as those corresponding to the distance between the Sr atoms belonging to the same $ab$-plane for the case of the constant $a^M_i$, and to the distance between the Sr atoms of the adjacent planes for the case of the constant $c^M_i$ (M indicates the shell M = STN or STO, and i = 1 ... n,) (see Fig.\ref{fig:structures}). In Tables \ref{tab:subcellsmultilayers} and \ref{tab:subcellsthinfilms} it can be observed that, even starting from the initial configuration where for each side of the interface the individual cell constants $a^M_i$ and $c^M_i$ are equal for all i within each M, the result after the optimization is more complex. Due to the periodic boundary conditions, the constants $a^M_i$ are equal for all i on both sides of the interface, obtaining an intermediate value of those corresponding to the STN and STO bulks. The obtained values are around 3.96 \angstrom in close agreement with the reported value in Ref. \cite{OKA2014} (3.98 \angstrom). Then the STN layer is compressed while the STO layer is elongated, resulting in a compression strain for the first and a tensile strain for the second, as expected. This gives rise to a possible ferroelectric behavior in both layers, as will be analyzed later. In contrast, the $c^M_i$ constants of the individual sub-cells, after structural relaxation result in a more complex scheme that will be analyzed in the next section. Fig. \ref{fig:PDOS}, shows the PDOS of the atoms for the SC V/4-SrTaO$_2$N/8-SrTiO$_3$ (V-4x8) case (the other structures show a similar behavior). Different shifts of the electronic levels compared with the pure compounds can be observed. These shifts are multiple due to the different value of the resulting $c^M_i$ constants and the local atomic rearrangement within each sub-cell. Previous studies on other heterostructures as CaZrO$_3$/SrTiO$_3$\cite{NAZIR2016, CHEN2015}, and AHfO$_3$/SrTiO$_3$ \cite{CHENG2016} show polarization induced and two-dimensional electron gas behavior, strongly dependent on strain, layer stacking and thickness. The insulator-to-metal transition was reported to occur at thickness strongly dependent on surface termination. In the present study, no bands crossing the Fermi energy are observed for any of the different heterojunctions studied, suggesting that the presence of nitrogen prevents this type of behavior.

{
\begin{table*}
\begin{center}
\begin{adjustbox}{width=0.9\textwidth}
\begin{tabular}{|c|c|c|c|c|c|c|c|c|c|c|c|c|c|c|}
\hline
 {\bf{Type}} & {\bf{Cell}} & $\delta \vp_{\mbox{{\tiny{ionic}}}}\atop{(10^{-3} C/m^2)} $ & ${a}\, \, (\mathrm{\mbox{{\tiny{\AA}}}} )$ & ${c}\, \,(\mathrm{\mbox{{\tiny{\AA}}}} )$ &  {\bf{Type}} & {\bf{Cell}} & $\delta \vp_{\mbox{{\tiny{ionic}}}}\atop{(10^{-3} C/m^2 )}$
  & ${a}\, \, (\mathrm{\mbox{{\tiny{\AA}}}} )$ & ${c}\, \,(\mathrm{\mbox{{\tiny{\AA}}}} )$
    & {\bf{Type}} &{\bf{Cell}} & $\delta \vp_{\mbox{{\tiny{ionic}}}}\atop{(10^{-3} C/m^2 )}$
  & ${a}\, \, (\mathrm{\mbox{{\tiny{\AA}}}} )$ & ${c}\, \,(\mathrm{\mbox{{\tiny{\AA}}}} )$
  \\ \hline
\multirow{10}{2em}{3x3}        &  &      &  \multirow{10}{2em}{3.960}    &  & \multirow{10}{2em}{4x4}        &  &      &  \multirow{10}{2em}{3.960}     &  & \multirow{10}{2em}{5x5}        & Ti 5 &  7.9    &  \multirow{10}{2em}{3.959}   & 3.871   \\ \cline{6-10}\cline{12-13} \cline{15-15}
      &  &      &     &  &       &  Ti 4 &    10.5  &       &  3.869 &  & Ti 4 &  2.8    &    & 3.942   \\
    \cline{1-5} \cline{7-8} \cline{10-10} \cline{12-13} \cline{15-15}
       & Ti 3 & 12.6     &      & 3.863 &         & Ti 3  &  3.5    &      & 3.935  &         & Ti 3 &  2.8    &     & 3.950  \\
  \cline{2-3} \cline{5-5} \cline{7-8} \cline{10-10} \cline{12-13} \cline{15-15}
        & Ti 2 &   4.4   &     &  3.926 &        &  Ti 2 &   3.3   &   & 3.941  &      & Ti 2 &  2.7    &     & 3.946   \\
  \cline{2-3} \cline{5-5} \cline{7-8} \cline{10-10} \cline{12-13} \cline{15-15}
         & Ti 1 &   4.1   &     &  3.906 &         & Ti 1 &  3.2   &    & 3.912 &       & Ti 1 &  2.6    &   & 3.917   \\
   \cline{2-3} \cline{5-5} \cline{7-8} \cline{10-10} \cline{12-13} \cline{15-15}
          & Ta 1 &    3.9  &      &  4.416 &         &  Ta 1 &   3.0   &     & 4.422 &        & Ta 1 &  2.4   &    & 4.430   \\
  \cline{2-3} \cline{5-5} \cline{7-8} \cline{10-10} \cline{12-13} \cline{15-15}
           &  Ta 2 &    12.1  &    & 4.324  &       &  Ta 2 &   9.2   &    & 4.326 &       & Ta 2 &  7.5    &   & 4.328 \\
  \cline{2-3} \cline{5-5} \cline{7-8} \cline{10-10} \cline{12-13} \cline{15-15}
       & Ta 3 &    12.5  &    &  4.354 &   & Ta 3 &    9.3  &   & 4.327 &    & Ta 3 &  7.5    &   & 4.326  \\
    \cline{1-5} \cline{7-8} \cline{10-10} \cline{12-13} \cline{15-15}
               &  &      &     &  &        &  Ta 4 &  9.5   &      &  4.362 &   & Ta 4 &  7.6    &    & 4.330  \\
  \cline{6-10} \cline{12-13} \cline{15-15}
              &  &      &    &  &      &  &      &      &  &       & Ta 5 &  7.97   &  & 4.365   \\   \hline
\end{tabular}
\end{adjustbox}
\caption{\label{tab:subcellsmultilayers}
Cell constants and ionic polarization obtained for the 3x3, 4x4 and 5x5 SCs after full relaxation of the atomic positions and cell constants. The numbering of the simple perovskite unit sub-cell starts with 1 just above/below the interface and increases as it moves up (for Ti sub-cells) or down (for Ta sub-cells) from it.}
\end{center}
\end{table*}
}

{
\begin{table*}
\begin{center}
\begin{adjustbox}{width=0.9\textwidth}
\begin{tabular}{|c|c|c|c|c|c|c|c|c|c|c|c|c|c|c|c|c|c|c|c|}
\hline
 {\bf{Type}} & {\bf{Cell}} & $\delta \vp_{\mbox{{\tiny{ionic}}}}\atop{{\mbox{\tiny{$(10^{-3} C/m^2 )$}}}} $ & ${a}\, \, (\mathrm{\mbox{{\tiny{\AA}}}} )$ & ${c}\, \,(\mathrm{\mbox{{\tiny{\AA}}}} )$ &  {\bf{Type}} & {\bf{Cell}} & $\delta \vp_{\mbox{{\tiny{ionic}}}}\atop{\mbox{\tiny{$(10^{-3} C/m^2 )$}}}$
  & ${a}\, \, (\mathrm{\mbox{{\tiny{\AA}}}} )$ & ${c}\, \,(\mathrm{\mbox{{\tiny{\AA}}}} )$
    & {\bf{Type}} &{\bf{Cell}} & $\delta \vp_{\mbox{{\tiny{ionic}}}}\atop{\mbox{\tiny{$(10^{-3} C/m^2 )$}}}$
  & ${a}\, \, (\mathrm{\mbox{{\tiny{\AA}}}} )$ & ${c}\, \,(\mathrm{\mbox{{\tiny{\AA}}}} )$  & {\bf{Type}} & {\bf{Cell}} & $\delta \vp_{\mbox{{\tiny{ionic}}}}\atop{\mbox{\tiny{$(10^{-3} C/m^2 )$}}}  $& ${a}\, \, (\mathrm{\mbox{{\tiny{\AA}}}} )$ & ${c}\, \,(\mathrm{\mbox{{\tiny{\AA}}}} )$
  \\ \hline
\multirow{15}{1em}{\rotatebox{90}{V-4x8 type SrO} }       &  &      &  \multirow{15}{2em}{3.954}    &  & \multirow{15}{1em}{\rotatebox{90}{V-4x8 type SrN} }       &  &      &  \multirow{15}{2em}{3.958}   & & \multirow{15}{1em}{\rotatebox{90}{V-5x10 type SrO} }       & Vac. &      &  \multirow{15}{2em}{3.956}    & 13.749 &\multirow{15}{1em}{\rotatebox{90}{V-5x10 type SrN}}        & Vac. &      &  \multirow{15}{2em}{3.959}    & 14.804 \\ \cline{12-13} \cline{15-15} \cline{17-18} \cline{20-20}
      &  &      &     &   &       &   &     &       &   &  & Ti 9 &  0.3    &    & 4.057 &  & Ti 9 & -0.2 & & 3.973 \\
    \cline{1-10} \cline{12-13} \cline{15-15} \cline{17-18} \cline{20-20}
      & Vac. &      &     & 13.738  &       &  Vac. &     &       &  13.923 &  & Ti 8 &  0.4    &    & 3.964 &  & Ti 8 & 0.0 & & 3.896 \\
    \cline{2-3} \cline{5-5} \cline{7-8} \cline{10-10} \cline{12-13} \cline{15-15} \cline{17-18} \cline{20-20}
        & Ti 7 &    0.2  &     & 4.060  &       &  Ti 7 & -0.2    &       &  4.062 &  & Ti 7 &  0.5    &    & 3.940 &  & Ti 7 & 0.1 & & 3.897 \\
    \cline{2-3} \cline{5-5} \cline{7-8} \cline{10-10} \cline{12-13} \cline{15-15} \cline{17-18} \cline{20-20}
          & Ti 6 &    0.4  &     & 3.960  &       &  Ti 6 &  0.1   &       &  3.961 &  & Ti 6 &  0.5    &    & 3.930 &  & Ti 6 & 0.1 & & 3.900 \\
    \cline{2-3} \cline{5-5} \cline{7-8} \cline{10-10} \cline{12-13} \cline{15-15} \cline{17-18} \cline{20-20}
          & Ti 5 &  0.4    &     & 3.937  &       &  Ti 5 & 0.1    &       &  3.934 &  & Ti 5 &  0.5    &    & 3.925 &  & Ti 5 & 0.0 & & 3.895 \\
    \cline{2-3} \cline{5-5} \cline{7-8} \cline{10-10} \cline{12-13} \cline{15-15} \cline{17-18} \cline{20-20}
          & Ti 4 &    0.4  &     & 3.925 &       &  Ti 4 &  0.1   &       &  3.923 &  & Ti 4 &  0.5    &    & 3.919 &  & Ti 4 & -0.0 & & 3.877 \\
    \cline{2-3} \cline{5-5} \cline{7-8} \cline{10-10} \cline{12-13} \cline{15-15} \cline{17-18} \cline{20-20}
          & Ti 3 &    0.4  &     & 3.921  &       &  Ti 3 & 0.1    &       &  3.917 &  & Ti 3 &  0.5    &    & 3.919 &  & Ti 3 & -0.1 & & 3.843 \\
    \cline{2-3} \cline{5-5} \cline{7-8} \cline{10-10} \cline{12-13} \cline{15-15} \cline{17-18} \cline{20-20}
          & Ti 2 &   0.4   &     & 3.911  &       &  Ti 2 & 0.1    &       &  3.897 &  & Ti 2 &  0.4    &    & 3.908 &  & Ti 2 & -0.1 & & 3.813 \\
    \cline{2-3} \cline{5-5} \cline{7-8} \cline{10-10} \cline{12-13} \cline{15-15} \cline{17-18} \cline{20-20}
          & Ti 1 &   0.3   &     & 3.863  &       &  Ti 1 &    -3.6  &       &  3.827 &  & Ti 1 &  0.4    &    &3.864 &  & Ti 1 & -3.2 & & 3.741 \\
    \cline{2-3} \cline{5-5} \cline{7-8} \cline{10-10} \cline{12-13} \cline{15-15} \cline{17-18} \cline{20-20}
          & Ta 1 &   -0.1   &     & 4.335  &       &  Ta 1 &  -3.3   &       &  4.290 &  & Ta 1 &  0.0    &    & 4.339 &  & Ta 1 & -2.9 & & 4.221 \\
    \cline{2-3} \cline{5-5} \cline{7-8} \cline{10-10} \cline{12-13} \cline{15-15} \cline{17-18} \cline{20-20}
          & Ta 2 &   3.6   &     & 4.245 &       & Ta 2&   -3.1  &       &  4.210 &  & Ta 2 &  3.1 &    & 4.248 &  & Ta 2 & -2.7 & & 4.154 \\
    \cline{2-3} \cline{5-5} \cline{7-8} \cline{10-10} \cline{12-13} \cline{15-15} \cline{17-18} \cline{20-20}
          & Ta 3 &   3.6   &     & 4.231 &       &  Ta 3 &  -3.1   &       &  4.194 &  & Ta 3 &  3.1    &    & 4.240 &  & Ta 3 & -2.7 & & 4.136 \\
    \cline{2-3} \cline{5-5} \cline{7-8} \cline{10-10} \cline{12-13} \cline{15-15} \cline{17-18} \cline{20-20}
          & Ta 4 &    3.4  &     & 4.150  &       &  Ta 4 & -0.0     &       &  4.133 &  & Ta 4 &  3.1    &    & 4.238 &  & Ta 4 & -2.7 & & 4.125 \\
    \cline{1-10} \cline{12-13} \cline{15-15} \cline{17-18} \cline{20-20}
              &  &      &    &  &      &  &      &      &  &       & Ta 5 &  2.8   &  & 4.143  & & Ta 5 & -0.1 & & 4.062\\   \hline
\end{tabular}
\end{adjustbox}
\caption{\label{tab:subcellsthinfilms} Cell constants and ionic polarization obtained for the V-4x8 and V-5x10 SCs types SrO and SrN, after full relaxation of the atomic positions and cell constants. The numbering of the simple perovskite unit sub-cell starts with 1 just above/below the interface and increases as it moves up (for Ti sub-cells) or down (for Ta sub-cells) from it. The vacuum layer is also included.}
\end{center}
\end{table*}
}

\section{Polarization analysis.}
\label{sec:polarization}
In this section the ionic and electronic contributions to the polarization are analyzed. For each SC, the ionic polarization of each sub-cell constituted by each individual perovskite cell structure was calculated.

It was not always clear how to calculate the polarization on a solid bulk, while studying a molecule or cluster with no net charge, it is defined straightforwardly as the separation of positive and negative charges in the system so that there is a net electric dipole moment per unit volume, where the ionic polarization is the one caused by relative displacements between positive and negative ions in ionic crystals.

With this definition in mind, and due to periodicity on bulk solids, the polarization on these systems is a multi-valued quantity, depending on the different choices of lattice vectors, that results in different ways of specifying the positions of the atoms. For each choice, a different polarization value is calculated, thus a lattice polarization can be defined instead of a polarization vector. If all the ions occupy positions with inversion symmetry, then that structure is non-polar.

Disagreements on how to calculate the polarization on solid bulks were ended with the so-called Modern Theory of Polarization\cite{KING1993}. If the shape and/or size of the unit cell are changed the polarization lattice will also change, as for example in response to strain. So, the meaning of an experimental measurement of the spontaneous polarization is the change of the polarization between an initial and final state. This theory provides the following formula in order to calculate the ionic contribution to the polarization:

\begin{equation}
\delta \vp_{\mbox{{\tiny{ionic}}}} = \vp^{f} - \vp^{\,0} = \frac{1}{V} \sum_{i} \left( q_{i}^{f} \ve{r}_{i}^{f} -q_{i}^{0} \ve{r}_{i}^{0} \right),
\label{eq:ionicpolarization}
\end{equation}

\noindent where $q_{i}$  stands for the ionic charge of the individual atom located at $\ve{r}_{i}$. The $f$ and $0$ upper indexes refer to the final (polarized) and the initial (normally non-polar) lattice structures. Considering symmetry provided by periodic conditions, the sum must be performed over non-equivalent atoms.

\subsection{Sub-cell ionic polarization.}
\label{subsec:ionicpol}

In the present study of SrTaO$_2$N/SrTiO$_3$ interfaces, in order to analyze the different contributions to the total polarization, the individual sub-cell ionic polarization is worth to be calculated. Still, care must be taken while applying the above stated formula. By construction, two different perovskite structures are matched together in order to generate the compound. Despite the whole SC is electrically neutral, this is not the case of each individual sub-cell. In particular, the sub-cells belonging to the interface have one N$^{-3}$ atom in one face and an O$^{-2}$ atom in the opposite face. After relaxation, due to tensile/constrain strain, each sub-cell ends up having a different volume: the same ai for all sub-cells, but different $c^{M}_{i}$.  This causes lack of periodicity in the $c$-direction. Moreover, the O, N and Ta/Ti atoms of the different octahedra move along the 001 direction different lengths. Thus, care must be taken for the individual sub-cell ionic polarization calculation due to the atoms on the lower $ab$-plane of each sub-cell are non-equivalent to those on the upper $ab$-plane. The ionic polarization of each sub-cell constituted by each individual perovskite cell structure was thus calculated as the difference between the dipole moments of each sub-cell, after and before displacement from the symmetric position, as stated in equation (\ref{eq:ionicpolarization}). Before relaxation, $\vp^{\,0}=0$ in almost all sub-cells, except those two with non-zero net charge.

First, the positions of both the net positive and negative charge centers in every single sub-cell are found. The net positive final charge center position $\ve{r}_{q^{+}}^{f}$ is calculated by a weighted sum of the positions $\ve{r}_{i, q^{+}}^{f}$ of each individual cation in the sub-cell:

\begin{equation}
\ve{r}_{q^{+}}^{f}= \frac{\sum_{i} m_i w_i \nu_i \ve{r}_{i, q^{+}}^{f}}{Q_{\mbox{\tiny{sub-cell}}}^{+}},
\label{eq:chargecenterposition}
\end{equation}

\noindent where the weight of each constituent is calculated by the multiplication of its valence charge $\nu_i$, the fraction $w_i$ (calculated by dividing the number of sub-cells that co-share this individual atom) and the multiplicity $m_i$ of equivalent atoms within the sub-cell, divided by the total positive charge of the sub-cell $Q_{\mbox{\tiny{sub-cell}}}^{+}$. A similar formula is used to calculate all the others charge position centers, replacing initial positions instead of final positions, and negative charge positions $\ve{r}_{i, q^{-}}^{f}$ and $Q_{\mbox{\tiny{sub-cell}}}^{-}$ for the anions.

A net electric dipole structure oriented on the $c$ direction is then formed, where the dipole displacement $\ve{d}^{f}= \ve{r}_{q^{+}}^{f}- \ve{r}_{q^{-}}^{f}$. Finally, the local ionic polarization is given by the dipole moment:

\begin{equation}
\vp^{f} = \frac{q_{e}}{V_{\mbox{\tiny{sub-cell}}}} \ve{d}^{f}.
\label{eq:dipolemoment}
\end{equation}

In the same way, $\vp^{\,0}_{\mbox{\tiny{ionic}}}$ is calculated. It must be remarked that there is no additive relation among the ionic polarization on individual sub-cells and the total ionic polarization of the whole SC structure. This is so, because in each individual sub-cell both upper and lower $ab$-planes atoms , as they are non-equivalents, were taken into account, so all atoms in the $ab$-planes contribute to two different sub-cells, whereas in the total ionic polarization each $ab$-planes atoms must be counted only once.

{\bf{-Multilayer n-SrTaO$_2$/n-SrTiO$_3$ ionic polarization.}}

\begin{figure*}
\begin{center}
\includegraphics[width=1\textwidth]{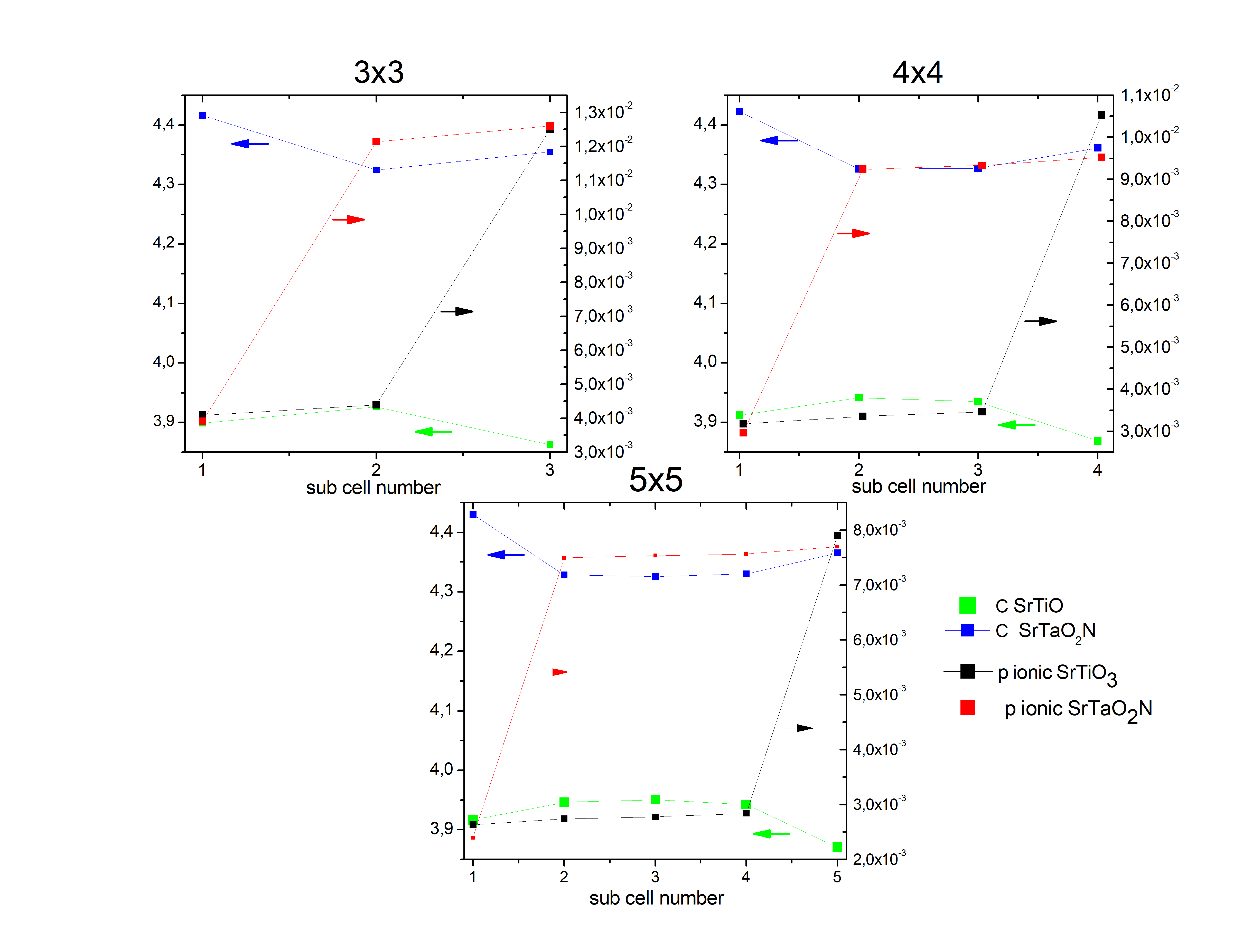}
\end{center}
\caption{Color online. Sub-cell lattice constant $c$ (left axis) and ionic polarization $\delta \vp$ (right axis) calculated for the 3x3, 4x4 and 5x5 SCs. The arrows and colors indicate the corresponding vertical axis.}
  \label{fig:multilayerspolarization}
\end{figure*}

In Table \ref{tab:subcellsmultilayers} the results obtained for the $c^M_i$ and the ionic polarization of each sub-cell in the systems n-SrTaO$_2$/n-SrTiO$_3$ (from here on ``nxn'') are presented. Each sub-cell has been labeled as Ta$_i$ and Ti$_i$ with $i=1...n$, being $i=1$ the sub-cell corresponding to the central interface. All these SCs were built with the first sub-cell having a symmetric O octahedron at the Ti side of the interface (i.e. corresponding to a TiO$_6$ octahedron) and a non-symmetric O-N octahedron at the Ta side (corresponding to a TaO$_5$N octahedron) for $i=1$. This geometry corresponds to the SrO termination-type. On the top and the bottom of the SC following the $c$ axis, periodic boundary conditions form other interfaces with the inverted symmetry. For example, the 3x3 SC has a TiO$_5$N (non-symmetric) and a TaO$_4$N$_2$ (symmetric trans-type) octahedron at each side of the interface for $i=3$. This corresponds to a SrN termination-type. In Table \ref{tab:subcellsmultilayers}, it can be observed that, for the Ti containing shells, the sub-cells present polarization values around $4.2 \times 10^{-3}$ C/m$^2$ for the 3x3, $3.3 \times 10^{-3}$ C/m$^2$ for the 4x4 and $2.7 \times 10^{-3}$ C/m$^2$ for the 5x5 in those with symmetric octahedra ($i=1 ... n-1$), and a bigger ionic polarization around three times the one obtained before ($12.1$, $10.5$ and $7.9 \times 10^{-3}$ C/m$^2$) at the non-symmetric sub-cell ($i=n$). In counterpart, for the Ta containing shell, the obtained polarization values are around $12.2 \times 10^{-3}$ C/m$^2$ for the 3x3, $9.3 \times 10^{-3}$ C/m$^2$ for the 4x4 and $7.5 \times 10^{-3}$ C/m$^2$ for the 5x5 for the symmetric trans-type sub-cells ($i= 2.... n$), with a decrease to around one third ($3.9$, $3.2$ and $2.4 \times 10^{-3}$ C/m$^2$ respectively) for the non-symmetric sub-cell ($i=1$). It looks like the non-symmetric sub-cell copies the polarization of the neighbor sub-cell on the opposite layer. This behavior can be observed in Fig. \ref{fig:multilayerspolarization}. In this Figure, the obtained values for the sub-cell lattice constant $c^M_i$ are depicted on the left axis and those corresponding to the ionic polarization on the right axis. It can be observed that in the three cases the sub-cell constant $c^{\mbox{\tiny{STN}}}_1$ results longer than the following in the same shell. This is due to the non-symmetry of the TaO$_5$N octahedron that produces an elongation of the corresponding sub-cell. On the opposite side, the last sub-cell in the Ti shell $c^{\mbox{\tiny{STO}}}_n$ results shorter than the previous in the same shell, in this case for the non-symmetric TiO$_5$N octahedron. In the figure it can also been observed an inverse relation between polarization and the sub-lattice constant $c^M_i$.

{\bf{-V/m-SrTaO$_2$N/n-SrTiO$_3$ ionic polarization.}}

\begin{figure*}
\begin{center}
\includegraphics[width = 1 \textwidth]{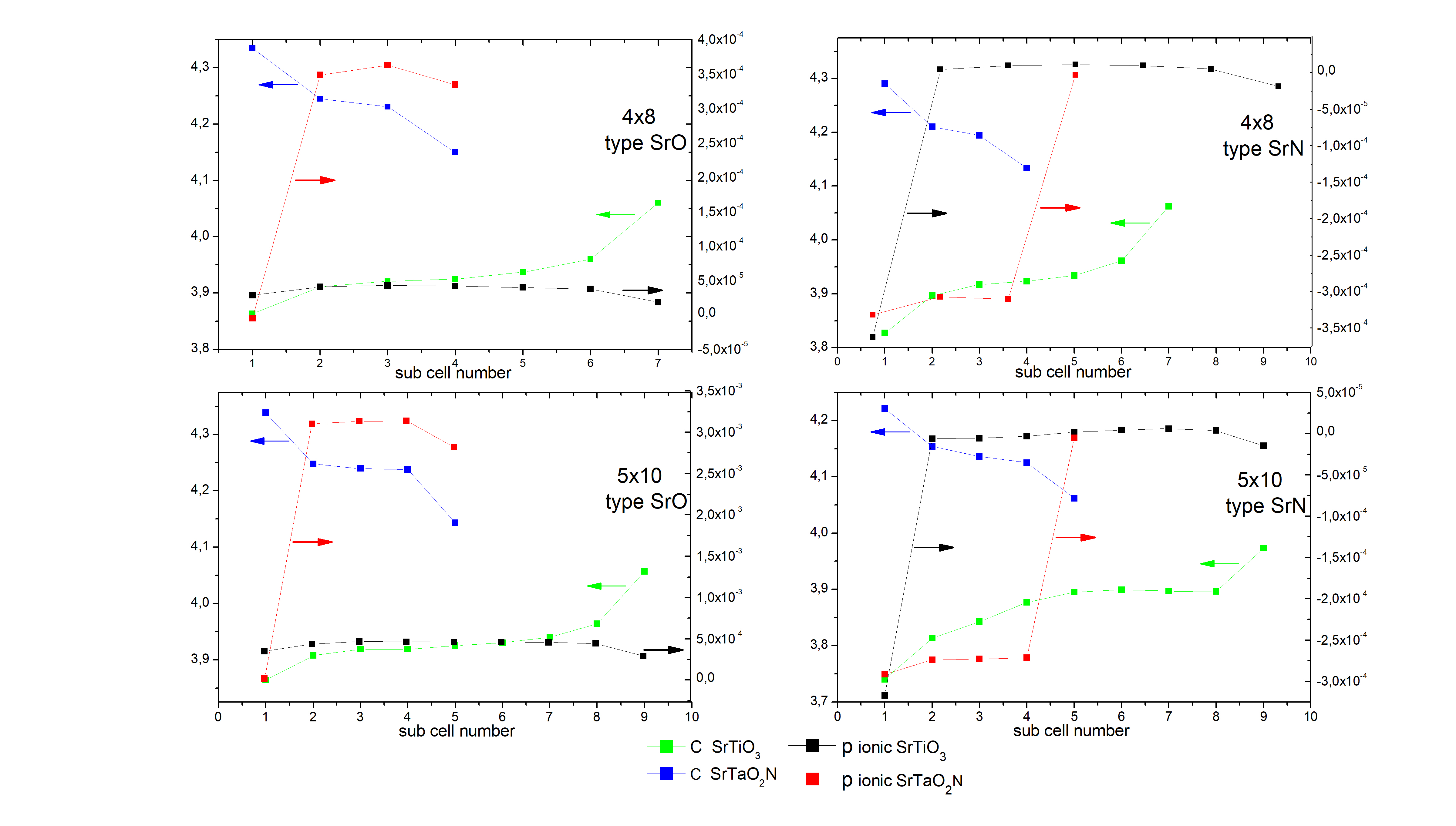}
\end{center}
\caption{Color online. Sub-cell lattice constant $c$ (left axis) and ionic polarization $\delta \vp$ (right axis)  for  the V-4x8 and V-5x10 (SrO and SrN-type) SCs.}
  \label{fig:thinfilmspolarization}
\end{figure*}

For the V/m-SrTaO$_2$N/n-SrTiO$_3$ SCs (from here on V-4x8 and V-5x10), two different interface terminations were built: one with the octahedra pair TiO$_6$-TaO$_5$N (SrO-type), and the other with the octahedra pair TiO$_5$N-TaO$_4$N$_2$ at the interface (SrN- type). Total energy considerations show that the SrO type are more stable than the SrN type ($\Delta$E = 7 meV and 11 meV for V-4x8 and V-5x10 respectively). In Table \ref{tab:subcellsthinfilms} and in Fig. \ref{fig:thinfilmspolarization} there are shown the results obtained for the $c^M_i$ and the ionic polarization of the respective internal sub-cells. The results show that the polarization on each sub-cell on both Ti and Ta shells are much lower than those obtained for the multilayer nxn structures. This effect may be associated with the fact that the electric dipole coupling for the nxn multilayer occurs from the top and the bottom faces of the SC, having the multilayers two (different) interfaces per SC, while for the V-mxn SCs there is only one coupling interface. Also, no large decrease in the ionic polarization as a function to the distance to the interface is observed, suggesting a collective behavior in each shell. The quasi-null value for the polarization for the last sub-cell Tan in the SrN interface type structure is due the lack of symmetry of the last TaO$_5$N octahedron. This fact is a requirement for charge neutrality considerations.

Comparing the ionic polarization between SrO and SrN interface-type structures, large differences can be observed: the ionic polarization resulted much lower in the Ti containing shells for SrN interface-type, due to the non-symmetric first TiO$_5$N octahedron.Also, the way to initialize the Ta layer affects the sign of the resulting polarization: in SrN interface-type cells, the Ta shells present polarization values with the opposite sign than that of the Ti shells, while for the SrO interface type, the individual sub-cell polarization are in the same direction. It should be noted that, the non-symmetric sub-cell seems to mimic the polarization of the symmetric sub-cell at the other side of the interface, as well as for the multilayers SCs type occurs. Another result to remark is that the sub-cell ionic polarization is lower for SrN interface-type comparing to the corresponding SrO one, and for the Ti containing shell for the SrN interface-type is nearly null for the V-5x10 SC.

\begin{figure*}
\begin{center}
\includegraphics[width=1\textwidth]{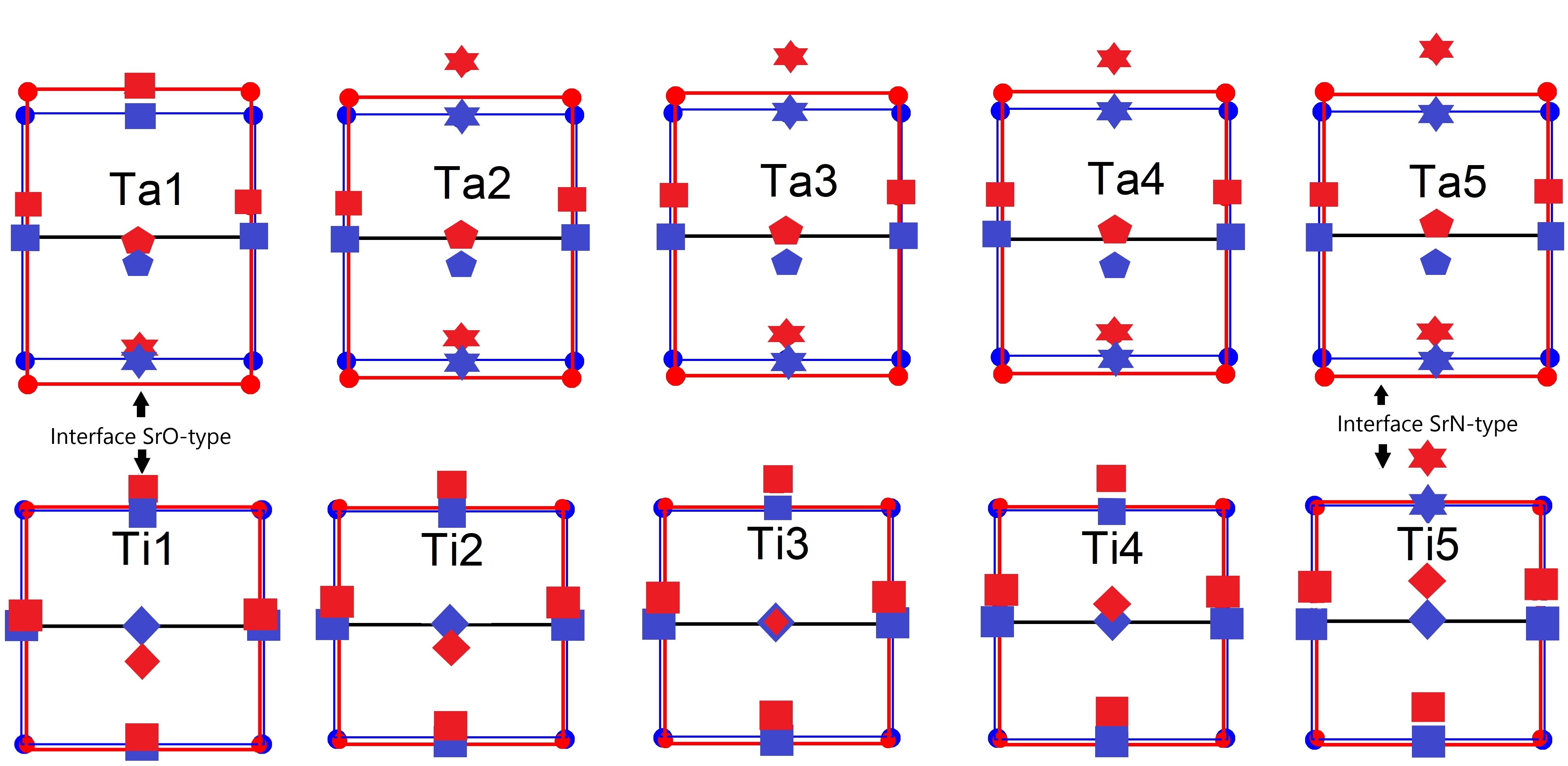}
\end{center}
\caption{Color online. Projection onto de $bc$-plane of the atomic positions and lattice constants before (blue) and after (red) atomic positions and cell relaxations in the 5x5 SC. The atomic symbols are Sr (circle), Ta (pentagon), O (square), N (star) and Ti (rhombus).}
  \label{fig:projectionatomposition}
\end{figure*}

	Another outstanding aspect in the analysis of the origin of the electric dipolar moment arises from the comparison of the individual ion displacement outside the central-symmetric positions. In the ferroelectric space group P4mm phases bulk ABO$_3$ perovskites, the dipole moment arises from the displacement in opposite directions of the central B cation and the O anions along the direction of the axis $c$ (in this representation, taking the A atoms position as the cell boundaries, and the $ab$-plane at $c/2$ as the symmetry plane). In contrast, for the present systems, a different behavior of the displacements of anions and cations is observed. Figure \ref{fig:projectionatomposition} shows a projection of the atomic positions on the $bc$-plane before and after the structural relaxation in the 5x5 multilayer SC. As mentioned, the starting structures were built with centrosymmetric positions for the Ti layers, and non-centrosymmetric positions for Ta layers, with a small initial displacement in one direction of the Ta cations (down from the central plane in Figure \ref{fig:projectionatomposition}). After the resulting atomic rearrangement due to relaxation, the Ta moves upward, now occupying positions slightly above the central symmetry plane, and the oxygen and nitrogen atoms move away in the same direction. This configuration results in a net dipole moment but based on different displacement lengths of cations and anions in the same direction. In the figure it can be observed that the only exception occurs in the first sub-cell of Ta corresponding to the interface, due to the asymmetry of the TaO$_5$N octahedron, in which the Ta remains slightly below the plane of symmetry, and the anions move above it. The individual sub-cell polarization at the Ti layer side present a similar behavior for some sub-cells (Ti 4-5 in Figure \ref{fig:projectionatomposition}) but not for all of them. Nevertheless, the local sub-cell polarization has the same sign in the five cases Ti 1-5. It should be mentioned that in order to ensure that the final atomic configurations have not fallen within a local energy minimum, different paths were taken to optimize the initial structures, resulting in similar final endpoints for all the pathways for each SC studied. In the Supplementary Information the whole atomic positions and cell parameter are given.

\subsection{Total Polarization.}
\label{subsec:totalpol}

Finally, the total (i.e. ionic plus electronic) polarization was computed for the optimized structures. For the calculations, both the Modern Theory of Polarization by means of the Berry Phase concept was used following the guidelines in the work of Spalding\cite{SPALDIN2012}, and also the Born's effective charge tensor were computed\cite{NEATON2005, ROY2010}. As in the previous subsection, for the calculation of $\delta \vp = \vp^{f} -\vp^{\,0}$, we defined the initial phase to that one built from the final relaxed structure by displacing the apical O/N atoms $z$ position in each octahedron to that of the corresponding Sr atoms plane, together with the displacement of the central Ta/Ti and O atoms $z$ position to the midpoint of the upper and lower Sr atoms plane. In this way we can vary continuously the atomic positions toward the final structure by means of a unique parameter $\lambda$ that runs from 0 to 1 and calculate the polarization variation as:

\begin{equation}
\delta \vp = \int_0^1 \frac{d \vp}{d \lambda} d \lambda.
\label{eq:lambdavariation}
\end{equation}

Within this method, the two types of contributions to the polarization are calculated separately:

\begin{equation}
\Delta \vP = \delta \vp_{\mbox{\tiny{ionic}}} +  \delta \vp_{\mbox{\tiny{electronic}}},
\label{eq:totalpol}
\end{equation}

\noindent and as a direct consequence, each component of the ionic and electronic polarization is defined except for a multiple of the corresponding component of the quantum of polarization:

\begin{equation}
\frac{q_e}{V_C} \vf{R} =\frac{q_e}{V_C} \vf{a}_1+ \vf{a}_2 + \vf{a}_3,
\label{eq:Bravaisvectors}
\end{equation}
                	
\noindent where $q_e$ is the electron charge, $V_C$ is the cell volume and the $\vf{a}_i$ are the vectors of the Bravais network of the structure associated with $V_C$.

\begin{figure}
\begin{center}
\includegraphics[width=0.5\textwidth]{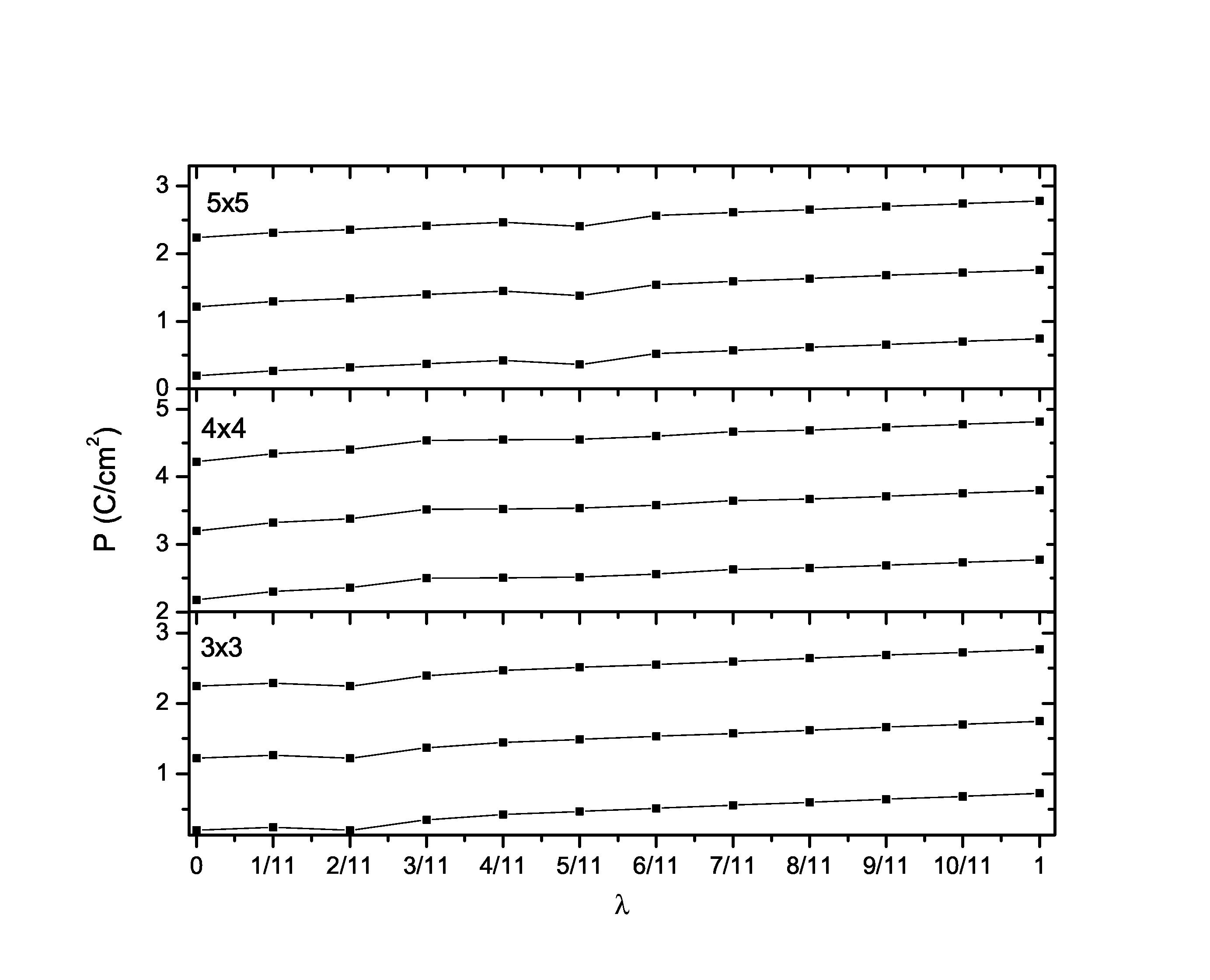}
\end{center}
\caption{Total Polarization as a function of the order parameter $\lambda$ for the 3x3, 4x4 and 5x5 SCs. The graphs show the increment of the calculated P by varying quasi continuously $\lambda$ form the initial structure ($\lambda=0$) to the final structure ($\lambda=1$). Three contiguous branches are plotted separated by the corresponding quantum of Polarization in the $c$-direction.}
  \label{fig:totalpolarization}
\end{figure}

Due to the above-mentioned multi-valued characteristic of the polarization, attention must be taken when going from one structure to the other in order to calculate the polarization variation $\Delta \vP$, in that the initial and final values correspond to the same branch. So, for each nxn SC, 10 intermediate structures were prepared, starting with the initial structure ($\lambda = 0$) and approaching gradually the final structure ($\lambda = 1$) by increasing $\lambda$ in $1/11$ steps. Also, polarization quanta were calculated for each structure. In this way we could identify and/or correct the calculated intermediate values in order to ensure they belong to the same branch. In Figure \ref{fig:totalpolarization} the obtained results are shown for three different branches for each nxn SC.
Born's effective charge is a tensor that relates the macroscopic polarization component in each $x_i$ direction with the collective displacement of a type-$k$ atom in the $x_j$ direction:

\begin{equation}
Z_{kij}=V \frac{\partial P_i}{\partial x_{kj}}\vert_{\vf{E}=0},
\label{eq:collectivedisplacement}
\end{equation}

where $V$ is the volume of the cell, and $\vf{E}$ is the macroscopic electric field. The effective Born's charge of a given atom is a dynamic load in the sense that represents the polarization response of a given type of atomic displacement:

\begin{equation}
\Delta P_i \cong \frac{1}{V} \sum_{k=1}^{N} \sum_{j=1}^{3} Z_{kij} \Delta x_{kj}.
\label{eq:polarizationwithdisplacement}
\end{equation}

In the last expression, $\Delta x_{kj}$ represents the atomic displacements from the initial to the final structures.

	The obtained results for  $\Delta \vP = \delta \vp_{\mbox{\tiny{ionic}}} +  \delta \vp_{\mbox{\tiny{electronic}}}$  in the 001 direction are presented in Table \ref{tab:totalpolarization}. For the 3x3, 4x4 and 5x5 multilayer structures, the polarization is around $-0.5$ C/m$^2$, while for the unique-interface structures V/4-SrTaO$_2$N/8-SrTiO$_3$ and V/5-SrTaO$_2$N/10-SrTiO$_3$ the corresponding values are lower: $\sim -0.029$ and $-0.039$ C/m$^2$ respectively. As a reference point, they should be compared with the corresponding to the well-known ferroelectric BaTiO$_3$, $\approx 0.26$ C/m$^2$. Also, it can be observed a very good agreement in the polarization values obtained by two very different calculation methods for the multilayer nxn structures. For the V/m-SrTaO$_2$N/n-SrTiO$_3$ structures the presence of the vacuum layer make the calculation of the polarization by the Berry phase method very difficult to obtain with a good convergence, so they were discarded and only those obtained by the Born's effective charges are shown. An additional advantage of the Born's effective charge calculation is that it provides the individual atomic contribution to the total polarization. The individual obtained values are included in the Supplementary Information. It is observed that, for the nxn multilayer structures,  the principal contribution arises from  the Ta and Ti with asymmetric octahedra (TaO$_5$N and TiO$_5$N), the Ti and Ta with symmetric TiO$_6$ and TiO$_4$N$_2$ at the interface, the N atoms, and the O atoms corresponding to the intermediate Ti octahedra (i.e., not belonging to an interface). On the other hand, for the V-4x8 and V-5x10 single interface structures, the mayor contributions come from the Ta atom at the interface, the single N at the vacuum surface, the O atoms at the interface, and the apical O atoms of the intermediate TiO$_6$ octahedra.
	
{
\begin{table}[h!]
\begin{center}
\begin{adjustbox}{width=0.5\textwidth}
\begin{tabular}{|c|c|c|}
\hline
 {\bf{Type of Super Cell}} & ${\mbox{\bf{Berry Phase Method}}}\atop{(C/m^2)} $& $ {\mbox{\bf{Born's Effective Charges Tensor Method}}}\atop{(C/m^2)} $ \\ \hline
 3x3 & -0.52 & -0.51 \\ \hline
 4x4 & -0.59 & -0.49 \\ \hline
 5x5 & -0.54 & -0.49 \\ \hline
 V-4x8 &  & -0.029 \\ \hline
 V-5x10 &  & -0.039 \\ \hline
\end{tabular}
\end{adjustbox}
\caption{\label{tab:totalpolarization}
  Total polarization for the different relaxed nxn and V-mxn SCs obtained using the Modern Theory of Polarization by the Berry phase method and the Born's effective charges tensor method. }
\end{center}
\end{table}
}	
	
Thus, this result is encouraging. The macroscopic polarization results similar for the three SCs under study. According to our knowledge this type of multilayer cells has not yet been fabricated. Although the difference in network constants between SrTiO$_3$ and SrTaO$_2$N is a bit large, the simple interface has been successfully constructed in the cited experimental works. This leads to expect that this type of periodic structure can also be manufactured. The results show that its development can be of great interest. In the case of theV/m-SrTaO$_2$N/n-SrTiO$_3$ structures, this last cell format is close to the experimental studies on thin film STN deposited on STO substrate. Although the value obtained for the total polarization is lower than in the case of multilayers, it was not zero, thus, the results presented in this work supports the experimental results presented in the work of D. Oka et al.\cite{OKA2014}.

\section{Summary and Conclusions.}
\label{sec:conclusions}

In this work we have analyzed the spontaneous rupture of inversion symmetry in two naturally dielectric compounds under normal conditions by the strain produced by the formation of a heterostructure. The determined net polarization appears as a consequence of the compression/tension stress that naturally arise between them under epitaxial growth due to the mismatch between their lattice constants. Different possible configurations were analyzed: the multilayer structure, and an approximation to a thin unique interface scheme. As a result of the proposed models, it was observed that the induced electrical polarization strongly depends on the termination layer at the interface, according to the content of oxygen or nitrogen atoms at the separation plane. The atomic displacement in each individual sub-cell was analyzed in detail, showing that the distortions that lead to electric polarization are not identical in each sub-cell, and that the displacement of anions and cations in the epitaxial direction does not follow the conventional scheme for phases with P4mm symmetry in perovskites. Finally, the total electric polarization was calculated, resulting for the multilayer structures in greater values than that of the reference compound BaTiO$_3$, leading to an open field to further experimental studies and a strong interest in the studied systems from the point of view of their possible industrial applications.

\section*{Acknowledgments.}

This work was partially supported by Consejo Nacional de Investigaciones Cient\'ificas y T\'ecnicas (CONICET) under Grants PIP11220170100987CO and PIO15520150100001CO, Facultad de Ingenier\'ia, Universidad Nacional de La Plata under Grant I191  and Facultad de Ciencias Exactas, Universidad Nacional de La Plata under Grant X843. We also thank the computational centers CSCAA, Aarhus Universitet, Denmark, and Proyecto Acelerado de C\'alculo of the SNCAD-MINCyT, Argentina. Also, to the Instituto de F\'isica La Plata-CONICET, Argentina, for the use of its facilities. Finally, to Dra. M. A. Taylor and L. A. Errico for careful reading and inspiring discussions.


\bibliographystyle{aipnum4-1}
\bibliography{SrTaO2NSrTiO3InterfacesarXiv}

\pagebreak
\section*{Supplementary Information.}
\setcounter{page}{1}
\renewcommand{\thepage}{S \arabic{page}}
\setcounter{table}{0}
\renewcommand{\thetable}{S\arabic{table}}
In this Supplementary Information, there are included seven tables (one for each SC in study) with the atomic positions, in fractional units, before and after relaxation, and the polarization contribution per atom, calculated by the Born's effective charge method.

{
\begin{table}[h!]
\begin{center}
\begin{adjustbox}{width=0.9\textwidth}
\begin{tabular}{|c|c|c|c|c|c|c|c|}
\hline
\multicolumn{8}{|c|}{3x3}\\ \hline						
\multirow{4}{1em}{\bf{A}}	&\multicolumn{3}{|c|}{\bf{CELL PARAMETERS (IN)}}	&	\multicolumn{3}{|c|}{\bf{CELL PARAMETERS (OUT)}} &	 \multirow{3}{8em}{ {\bf{Born's Effective Charge Method}}} 	\\ \cline{2-7}
& ${a}\, \, (\mathrm{\mbox{{\tiny{\AA}}}} )$ & ${b}\, \,(\mathrm{\mbox{{\tiny{\AA}}}} )$ & ${c}\, \, (\mathrm{\mbox{{\tiny{\AA}}}} )$ & ${a}\, \,(\mathrm{\mbox{{\tiny{\AA}}}} )$ & ${b}\, \, (\mathrm{\mbox{{\tiny{\AA}}}} )$ & ${c}\, \,(\mathrm{\mbox{{\tiny{\AA}}}} )$ &  \\ \cline{2-7}
	&	3.969	& 3.969 & 	23.838298218 & 	3.960428929	& 3.960428929 & 24.789206204	 & \\ \cline{2-7}
&	\multicolumn{3}{|c|}{\bf{ATOMIC POSITIONS (crystal)}} & \multicolumn{3}{|c|}{\bf{ATOMIC POSITIONS (crystal)}}	&	${(C/m^2)} $	\\ \hline
Sr & 0.0	& 0.0	& 0.50856396	& 0.0   & 0.0 	& 0.512883519	& 0.0000 \\ \hline
Sr	& 0.0	& 0.0	& 0.67237597	& 0.0	& 0.0	& 0.670470348	& 0.0000 \\ \hline
Sr	& 0.0	& 0.0	& 0.83618799	& 0.0	& 0.0	& 0.828852174	& 0.0000 \\ \hline
Sr	& 0.0	& 0.0	& 0.00000000	& 0.0	& 0.0	& -0.015322074	& 0.0000 \\ \hline
Sr	& 0.0	& 0.0	& 0.16952132	& 0.0	& 0.0	& 0.160333568	& 0.0000 \\ \hline
Sr	& 0.0	& 0.0	& 0.33904264	& 0.0	& 0.0	& 0.334730332	& 0.0000 \\ \hline
Ta	& 0.5	& 0.5	& 0.24607789	& 0.5	& 0.5	& 0.074220775	& 0.0117 \\ \hline
Ta	& 0.5	& 0.5	& 0.41559921	& 0.5	& 0.5	& 0.247837171	& 0.0020 \\ \hline
Ta	& 0.5	& 0.5	& 0.59046997	& 0.5	& 0.5	& 0.421397456	& -0.0207 \\ \hline
O	& 0.0	& 0.5	& 0.59046997	& 0.0	& 0.5	& 0.594217841	& -0.0071 \\ \hline
O	& 0.5	& 0.0	& 0.59046997	& 0.5	& 0.0	& 0.594217841	& -0.0071 \\ \hline
O	& 0.0	& 0.5	& 0.75428198	& 0.0	& 0.5	& 0.754028083	& -0.0124 \\ \hline
O	& 0.5	& 0.0	& 0.75428198	& 0.5	& 0.0	& 0.754028083	& -0.0124 \\ \hline
O	& 0.0	& 0.5	& 0.91809399	& 0.0	& 0.5	& 0.913686997	& 0.0020 \\ \hline
O	& 0.5	& 0.0	& 0.91809399	& 0.5	& 0.0	& 0.913686997	& 0.0020 \\ \hline
O	& 0.0	& 0.5	& 0.08556107	& 0.0	& 0.5	& 0.086148341	& -0.0213 \\ \hline
O	& 0.5	& 0.0	& 0.08556107	& 0.5	& 0.0	& 0.086148341	& -0.0213 \\ \hline
O	& 0.0	& 0.5	& 0.25508239	& 0.0	& 0.5	& 0.259777265	& -0.0189 \\ \hline
O	& 0.5	& 0.0	& 0.25508239	& 0.5	& 0.0	& 0.259777265	& -0.0189 \\ \hline
O	& 0.0	& 0.5	& 0.42460371	& 0.0	& 0.5	& 0.434138139	& -0.0217 \\ \hline
O	& 0.5	& 0.0	& 0.42460371	& 0.5	& 0.0	& 0.434138139	& -0.0217 \\ \hline
O	& 0.0	& 0.5	& 0.50856396	& 0.0	& 0.5	& 0.514815262	& -0.0160 \\ \hline
O	& 0.5	& 0.5	& 0.67237597	& 0.5	& 0.5	& 0.674414641	& -0.0319 \\ \hline
O	& 0.5	& 0.5	& 0.83618799	& 0.5	& 0.5	& 0.834009673	& -0.0384 \\ \hline
N	& 0.5	& 0.5	& 0.00054814	& 0.5	& 0.5	& -0.001170743	& -0.0847 \\ \hline
N	& 0.5	& 0.5	& 0.17006945	& 0.5	& 0.5	& 0.172573261	& -0.0587 \\ \hline
N	& 0.5	& 0.5	& 0.33959077	& 0.5	& 0.5	& 0.346412181	& -0.0596 \\ \hline
Ti	& 0.5	& 0.5	& 0.59046997	& 0.5	& 0.5	& 0.589679377	& -0.0206 \\ \hline
Ti	& 0.5	& 0.5	& 0.75428198	& 0.5	& 0.5	& 0.74984407	    & 0.0018 \\ \hline
Ti	& 0.5	& 0.5	& 0.91809399	& 0.5	& 0.5	& 0.910319667	& -0.0386 \\ \hline
\end{tabular}
\end{adjustbox}
\caption{Atomic positions and polarization contributions per atom in the 3x3 SC. }
\end{center}
\end{table}

{
\begin{table}
\begin{center}
\begin{adjustbox}{width=1\textwidth}
\begin{tabular}{|c|c|c|c|c|c|c|c|}
\hline
\multicolumn{8}{|c|}{4x4}\\ \hline						
\multirow{4}{1em}{\bf{A}}	&\multicolumn{3}{|c|}{\bf{CELL PARAMETERS (IN)}}	&	\multicolumn{3}{|c|}{\bf{CELL PARAMETERS (OUT)}} &	 \multirow{3}{8em}{ {\bf{Born's Effective Charge Method}}} 	\\ \cline{2-7}
& ${a}\, \, (\mathrm{\mbox{{\tiny{\AA}}}} )$ & ${b}\, \,(\mathrm{\mbox{{\tiny{\AA}}}} )$ & ${c}\, \, (\mathrm{\mbox{{\tiny{\AA}}}} )$ & ${a}\, \,(\mathrm{\mbox{{\tiny{\AA}}}} )$ & ${b}\, \, (\mathrm{\mbox{{\tiny{\AA}}}} )$ & ${c}\, \,(\mathrm{\mbox{{\tiny{\AA}}}} )$ &  \\ \cline{2-7}
	&	3.969	& 3.969 & 	31.784398682 & 	3.956501132	& 3.956501132 & 33.0941638732	 & \\ \cline{2-7}
&	\multicolumn{3}{|c|}{\bf{ATOMIC POSITIONS (crystal)}} & \multicolumn{3}{|c|}{\bf{ATOMIC POSITIONS (crystal)}}	&	${(C/m^2)} $	\\ \hline
Sr	& 0.0	& 0.0	& 0.50856397	& 0.0	& 0.0	& 0.513492499	& 0.0000 \\ \hline
Sr	& 0.0	& 0.0	& 0.63142298	& 0.0	& 0.0	& 0.631696332	& 0.0000\\ \hline
Sr	& 0.0	& 0.0	& 0.75428198	& 0.0	& 0.0	& 0.750793336	& 0.0000\\ \hline
Sr	& 0.0	& 0.0	& 0.87714099	& 0.0	& 0.0	& 0.869695092	& 0.0000\\ \hline
Sr	& 0.0	& 0.0	& 0.00000000	& 0.0	& 0.0	& -0.013400303	& 0.0000\\ \hline
Sr	& 0.0	& 0.0	& 0.12714098	& 0.0	& 0.0	& 0.118396546	& 0.0000\\ \hline
Sr	& 0.0	& 0.0	& 0.25428197	& 0.0	& 0.0	& 0.249142908	& 0.0000\\ \hline
Sr	& 0.0	& 0.0	& 0.38142295	& 0.0	& 0.0	& 0.379870888	& 0.0000\\ \hline
Ta	& 0.5	& 0.5	& 0.05741742	& 0.5	& 0.5	& 0.053812858	& 0.0121\\ \hline
Ta	& 0.5	& 0.5	& 0.18455841	& 0.5	& 0.5	& 0.184211677	& 0.0031\\ \hline
Ta	& 0.5	& 0.5	& 0.31169939	& 0.5	& 0.5	& 0.314807561	& 0.0015\\ \hline
Ta	& 0.5	& 0.5	& 0.43884038	& 0.5	& 0.5	& 0.444961055	& -0.0044\\ \hline
O	& 0.5	& 0.0	& 0.56999347	& 0.5	& 0.0	& 0.574605083	& -0.0050\\ \hline
O	& 0.0	& 0.5	& 0.56999347	& 0.0	& 0.5	& 0.574605083	& -0.0050\\ \hline
O	& 0.5	& 0.0	& 0.69285248	& 0.5	& 0.0	& 0.694279006	& -0.0118\\ \hline
O	& 0.0	& 0.5	& 0.69285248	& 0.0	& 0.5	& 0.694279006	& -0.0118\\ \hline
O	& 0.5	& 0.0	& 0.81571149	& 0.5	& 0.0	& 0.813854193	& -0.0157\\ \hline
O	& 0.0	& 0.5	& 0.81571149	& 0.0	& 0.5	& 0.813854193	& -0.0157\\ \hline
O	& 0.5	& 0.0	& 0.93857049	& 0.5	& 0.0	& 0.933523277	& 0.0022\\ \hline
O	& 0.0	& 0.5	& 0.93857049	& 0.0	& 0.5	& 0.933523277	& 0.0022\\ \hline
O	& 0.5	& 0.0	& 0.0641708	& 0.5	& 0.0	& 0.062890568	& -0.0176\\ \hline
O	& 0.0	& 0.5	& 0.0641708	& 0.0	& 0.5	& 0.062890568	& -0.0176\\ \hline
O	& 0.5	& 0.0	& 0.19131178	& 0.5	& 0.0	& 0.193277437	& -0.0148\\ \hline
O	& 0.0	& 0.5	& 0.19131178	& 0.0	& 0.5	& 0.193277437	& -0.0148\\ \hline
O	& 0.5	& 0.0	& 0.31845276	& 0.5	& 0.0	& 0.323866846	& -0.0101\\ \hline
O	& 0.0	& 0.5	& 0.31845276	& 0.0	& 0.5	& 0.323866846	& -0.0101\\ \hline
O	& 0.5	& 0.0	& 0.44559375	& 0.5	& 0.0	& 0.454605267	& -0.0096\\ \hline
O	& 0.0	& 0.5	& 0.44559375	& 0.0	& 0.5	& 0.454605267	& -0.0096\\ \hline
O	& 0.0	& 0.5	& 0.50856397	& 0.5	& 0.5	& 0.515118015	& -0.0074\\ \hline
O	& 0.5	& 0.5	& 0.63142298	& 0.5	& 0.5	& 0.634650822	& -0.0368\\ \hline
O	& 0.5	& 0.5	& 0.75428198	& 0.5	& 0.5	& 0.754161882	& -0.0419\\ \hline
O	& 0.5	& 0.5	& 0.87714099	& 0.5	& 0.5	& 0.873803707	& -0.0279\\ \hline
N	& 0.5	& 0.5	& 0.0004111	& 0.5	& 0.5	& -0.002607276	& -0.0604\\ \hline
N	& 0.5	& 0.5	& 0.12755208	& 0.5	& 0.5	& 0.127862839	& -0.0524\\ \hline
N	& 0.5	& 0.5	& 0.25469307	& 0.5	& 0.5	& 0.258490199	& -0.0428\\ \hline
N	& 0.5	& 0.5	& 0.38183405	& 0.5	& 0.5	& 0.388841466	& -0.0236\\ \hline
Ti & 0.5	& 0.5	& 0.56999347	& 0.5	& 0.5	& 0.57107449		& -0.0172\\ \hline
Ti & 0.5	& 0.5	& 0.69285248	& 0.5	& 0.5	& 0.69092418		& -0.0054\\ \hline
Ti & 0.5	& 0.5	& 0.81571149	& 0.5	& 0.5	& 0.810629077	& 0.0053\\ \hline
Ti & 0.5	& 0.5	& 0.93857049	& 0.5	& 0.5	& 0.930880407	& -0.0210\\ \hline
\end{tabular}
\end{adjustbox}
\caption{Atomic positions and polarization contributions per atom in the 4x4 SC. }
\end{center}
\end{table}

{
\begin{table}
\begin{center}
\begin{adjustbox}{width=0.9\textwidth}
\begin{tabular}{|c|c|c|c|c|c|c|c|}
\hline
\multicolumn{8}{|c|}{5x5}\\ \hline						
\multirow{4}{1em}{\bf{A}}	&\multicolumn{3}{|c|}{\bf{CELL PARAMETERS (IN)}}	&	\multicolumn{3}{|c|}{\bf{CELL PARAMETERS (OUT)}} &	 \multirow{3}{8em}{ {\bf{Born's Effective Charge Method}}} 	\\ \cline{2-7}
& ${a}\, \, (\mathrm{\mbox{{\tiny{\AA}}}} )$ & ${b}\, \,(\mathrm{\mbox{{\tiny{\AA}}}} )$ & ${c}\, \, (\mathrm{\mbox{{\tiny{\AA}}}} )$ & ${a}\, \,(\mathrm{\mbox{{\tiny{\AA}}}} )$ & ${b}\, \, (\mathrm{\mbox{{\tiny{\AA}}}} )$ & ${c}\, \,(\mathrm{\mbox{{\tiny{\AA}}}} )$ &  \\ \cline{2-7}
	&	3.969	& 3.969 & 	39.730497032 & 	3.9587355115	& 3.9587355115 & 41.404531081	 & \\ \cline{2-7}
&	\multicolumn{3}{|c|}{\bf{ATOMIC POSITIONS (crystal)}} & \multicolumn{3}{|c|}{\bf{ATOMIC POSITIONS (crystal)}}	&	${(C/m^2)} $	\\ \hline
Sr	& 0.0	& 0.0	& 0.50856393	& 0.0	& 0.0	& 0.513859138	& 0.0000\\ \hline
Sr	& 0.0	& 0.0	& 0.60685115	& 0.0	& 0.0	& 0.608449083	& 0.0000\\ \hline
Sr	& 0.0	& 0.0	& 0.70513835	& 0.0	& 0.0	& 0.703753505	& 0.0000\\ \hline
Sr	& 0.0	& 0.0	& 0.80342556	& 0.0	& 0.0	& 0.799159185	& 0.0000\\ \hline
Sr	& 0.0	& 0.0	& 0.90171276	& 0.0	& 0.0	& 0.894371574	& 0.0000\\ \hline
Sr	& 0.0	& 0.0	& 0.00000000	& 0.0	& 0.0	& -0.012153008	& 0.0000\\ \hline
Sr	& 0.0	& 0.0	& 0.10171279	& 0.0	& 0.0	& 0.093283716	& 0.0000\\ \hline
Sr	& 0.0	& 0.0	& 0.20342557	& 0.0	& 0.0	& 0.197858056	& 0.0000\\ \hline
Sr	& 0.0	& 0.0	& 0.30513836	& 0.0	& 0.0	& 0.302329211	& 0.0000\\ \hline
Sr	& 0.0	& 0.0	& 0.40685115	& 0.0	& 0.0	& 0.406858615	& 0.0000\\ \hline
Ta	& 0.5	& 0.5	& 0.04593394	& 0.5	& 0.5	& 0.041631719	& 0.0053\\ \hline
Ta	& 0.5	& 0.5	& 0.14764672	& 0.5	& 0.5	& 0.145929734	& 0.0014\\ \hline
Ta	& 0.5	& 0.5	& 0.24935951	& 0.5	& 0.5	& 0.250497263	& 0.0016\\ \hline
Ta	& 0.5	& 0.5	& 0.3510723	& 0.5	& 0.5	& 0.354925433	& 0.0017\\ \hline
Ta	& 0.5	& 0.5	& 0.45278509	& 0.5	& 0.5	& 0.459073567	& -0.0118\\ \hline
O	& 0.5	& 0.0	& 0.55770754	& 0.5	& 0.0	& 0.562846096	& -0.0061\\ \hline
O	& 0.0	& 0.5	& 0.55770754	& 0.0	& 0.5	& 0.562846096	& -0.0061\\ \hline
O	& 0.5	& 0.0	& 0.65599474	& 0.5	& 0.0	& 0.658534603	& -0.0088\\ \hline
O	& 0.0	& 0.5	& 0.65599474	& 0.0	& 0.5	& 0.658534603	& -0.0088\\ \hline
O	& 0.5	& 0.0	& 0.75428195	& 0.5	& 0.0	& 0.7540433		& -0.0093\\ \hline
O	& 0.0	& 0.5	& 0.75428195	& 0.0	& 0.5	& 0.7540433		& -0.0093\\ \hline
O	& 0.5	& 0.0	& 0.85256917	& 0.5	& 0.0	& 0.849728716	& -0.0110\\ \hline
O	& 0.0	& 0.5	& 0.85256917	& 0.0	& 0.5	& 0.849728716	& -0.0110\\ \hline
O	& 0.5	& 0.0	& 0.95085636	& 0.5	& 0.0	& 0.945447236	& 0.0058\\ \hline
O	& 0.0	& 0.5	& 0.95085636	& 0.0	& 0.5	& 0.945447236	& 0.0058\\ \hline
O	& 0.5	& 0.0	& 0.05133664	& 0.5	& 0.0	& 0.04897692		& -0.0091\\ \hline
O	& 0.0	& 0.5	& 0.05133664	& 0.0	& 0.5	& 0.04897692		& -0.0091\\ \hline
O	& 0.5	& 0.0	& 0.15304942	& 0.5	& 0.0	& 0.153259942	& -0.0068\\ \hline
O	& 0.0	& 0.5	& 0.15304942	& 0.0	& 0.5	& 0.153259942	& -0.0068\\ \hline
O	& 0.5	& 0.0	& 0.25476221	& 0.5	& 0.0	& 0.257813927	& -0.0068\\ \hline
O	& 0.0	& 0.5	& 0.25476221	& 0.0	& 0.5	& 0.257813927	& -0.0068\\ \hline
O	& 0.5	& 0.0	& 0.356475		& 0.5	& 0.0	& 0.362244325	& -0.0090\\ \hline
O	& 0.0	& 0.5	& 0.356475		& 0.0	& 0.5	& 0.362244325	& -0.0090\\ \hline
O	& 0.5	& 0.0	& 0.45818779	& 0.5	& 0.0	& 0.46688181		& -0.0150\\ \hline
O	& 0.0	& 0.5	& 0.45818779	& 0.0	& 0.5	& 0.46688181		& -0.0150\\ \hline
O	& 0.5	& 0.5	& 0.50856393	& 0.5	& 0.5	& 0.515243343	& -0.0135\\ \hline
O	& 0.5	& 0.5	& 0.60685115	& 0.5	& 0.5	& 0.610859859	& -0.0253\\ \hline
O	& 0.5	& 0.5	& 0.70513835	& 0.5	& 0.5	& 0.706363373	& -0.0264\\ \hline
O	& 0.5	& 0.5	& 0.80342556	& 0.5	& 0.5	& 0.801972698	& -0.0285\\ \hline
O	& 0.5	& 0.5	& 0.90171276	& 0.5	& 0.5	& 0.897663573	& -0.0299\\ \hline
N	& 0.5	& 0.5	& 0.00032888	& 0.5	& 0.5	& -0.003439397	& -0.0477\\ \hline
N	& 0.5	& 0.5	& 0.10204167	& 0.5	& 0.5	& 0.100924515	& -0.0246\\ \hline
N	& 0.5	& 0.5	& 0.20375445	& 0.5	& 0.5	& 0.205492647	& -0.0214\\ \hline
N	& 0.5	& 0.5	& 0.30546724	& 0.5	& 0.5	& 0.309934728	& -0.0237\\ \hline
N	& 0.5	& 0.5	& 0.40718003	& 0.5	& 0.5	& 0.414233145	& -0.0328\\ \hline
Ti	& 0.5	& 0.5	& 0.55770754	& 0.5	& 0.5	& 0.559925839	& -0.0163\\ \hline
Ti	& 0.5	& 0.5	& 0.65599474	& 0.5	& 0.5	& 0.655760579	& -0.0045\\ \hline
Ti	& 0.5	& 0.5	& 0.75428195	& 0.5	& 0.5	& 0.751320378	& -0.0018\\ \hline
Ti	& 0.5	& 0.5	& 0.85256917	& 0.5	& 0.5	& 0.84707003		& 0.0040\\ \hline
Ti	& 0.5	& 0.5	& 0.95085636	& 0.5	& 0.5	& 0.943226749	& -0.0444\\ \hline
\end{tabular}
\end{adjustbox}
\caption{Atomic positions and polarization contributions per atom in the 5x5 SC. }
\end{center}
\end{table}
}

{
\begin{table}
\begin{center}
\begin{adjustbox}{width=0.75\textwidth}
\begin{tabular}{|c|c|c|c|c|c|c|c|}
\hline
\multicolumn{8}{|c|}{V-4x8 Type SrO}\\ \hline						
\multirow{4}{1em}{\bf{A}}	&\multicolumn{3}{|c|}{\bf{CELL PARAMETERS (IN)}}	&	\multicolumn{3}{|c|}{\bf{CELL PARAMETERS (OUT)}} &	 \multirow{3}{8em}{ {\bf{Born's Effective Charge Method}}} 	\\ \cline{2-7}
& ${a}\, \, (\mathrm{\mbox{{\tiny{\AA}}}} )$ & ${b}\, \,(\mathrm{\mbox{{\tiny{\AA}}}} )$ & ${c}\, \, (\mathrm{\mbox{{\tiny{\AA}}}} )$ & ${a}\, \,(\mathrm{\mbox{{\tiny{\AA}}}} )$ & ${b}\, \, (\mathrm{\mbox{{\tiny{\AA}}}} )$ & ${c}\, \,(\mathrm{\mbox{{\tiny{\AA}}}} )$ &  \\ \cline{2-7}
	&	3.969	& 3.969 & 	57.404396876 & 	3.9542	& 3.9542 & 58.2759	 & \\ \cline{2-7}
&	\multicolumn{3}{|c|}{\bf{ATOMIC POSITIONS (crystal)}} & \multicolumn{3}{|c|}{\bf{ATOMIC POSITIONS (crystal)}}	&	${(C/m^2)} $	\\ \hline
Sr	& 0.0	& 0.0	& 0.28158819	& 0.0	& 0.0	& 0.285554998	& 0.0000\\ \hline
Sr	& 0.0	& 0.0	& 0.34961433	& 0.0	& 0.0	& 0.351845779	& 0.0000\\ \hline
Sr	& 0.0	& 0.0	& 0.41764047	& 0.0	& 0.0	& 0.418960923	& 0.0000\\ \hline
Sr	& 0.0	& 0.0	& 0.48566661	& 0.0	& 0.0	& 0.486249907	& 0.0000\\ \hline
Sr	& 0.0	& 0.0	& 0.55369276	& 0.0	& 0.0	& 0.553604311	& 0.0000\\ \hline
Sr	& 0.0	& 0.0	& 0.62171889	& 0.0	& 0.0	& 0.621153847	& 0.0000\\ \hline
Sr	& 0.0	& 0.0	& 0.68974504	& 0.0	& 0.0	& 0.689105589	& 0.0000\\ \hline
Sr	& 0.0	& 0.0	& 0.75777117	& 0.0	& 0.0	& 0.758781586	& 0.0000\\ \hline
Sr	& 0.0	& 0.0	& 0.00000000	& 0.0	& 0.0	& -0.005480721	& 0.0000\\ \hline
Sr	& 0.0	& 0.0	& 0.07039704	& 0.0	& 0.0	& 0.065729618	& 0.0000\\ \hline
Sr	& 0.0	& 0.0	& 0.14079408	& 0.0	& 0.0	& 0.138334888	& 0.0000\\ \hline
Sr	& 0.0	& 0.0	& 0.21119112	& 0.0	& 0.0	& 0.211174709	& 0.0000\\ \hline
Ta	& 0.5	& 0.5	& 0.03179161	& 0.5	& 0.5	& 0.028001869	& -0.0005\\ \hline
Ta	& 0.5	& 0.5	& 0.10218865	& 0.5	& 0.5	& 0.101256077	& 0.0001\\ \hline
Ta	& 0.5	& 0.5	& 0.17258569	& 0.5	& 0.5	& 0.173939898	& -0.0001\\ \hline
Ta	& 0.5	& 0.5	& 0.24298273	& 0.5	& 0.5	& 0.24634435	& -0.0027\\ \hline
O	& 0.5	& 0.0	& 0.79178425	& 0.5	& 0.0	& 0.790075398	& -0.0002\\ \hline
O	& 0.0	& 0.5	& 0.79178425	& 0.0	& 0.5	& 0.790075398	& -0.0002\\ \hline
O	& 0.5	& 0.0	& 0.7237581	    & 0.5	& 0.0	& 0.722653675	& 0.0000\\ \hline
O	& 0.0	& 0.5	& 0.7237581	    & 0.0	& 0.5	& 0.722653675	& 0.0000\\ \hline
O	& 0.5	& 0.0	& 0.65573197	& 0.5	& 0.0	& 0.655126097	& -0.0002\\ \hline
O	& 0.0	& 0.5	& 0.65573197	& 0.0	& 0.5	& 0.655126097	& -0.0002\\ \hline
O	& 0.5	& 0.0	& 0.58770583	& 0.5	& 0.0	& 0.587589422	& -0.0008\\ \hline
O	& 0.0	& 0.5	& 0.58770583	& 0.0	& 0.5	& 0.587589422	& -0.0008\\ \hline
O	& 0.5	& 0.0	& 0.51967968	& 0.5	& 0.0	& 0.520214145	& 0.0006\\ \hline
O	& 0.0	& 0.5	& 0.51967968	& 0.0	& 0.5	& 0.520214145	& 0.0006\\ \hline
O	& 0.5	& 0.0	& 0.45165354	& 0.5	& 0.0	& 0.452909777	& -0.0002\\ \hline
O	& 0.0	& 0.5	& 0.45165354	& 0.0	& 0.5	& 0.452909777	& -0.0002\\ \hline
O	& 0.5	& 0.0	& 0.38362741	& 0.5	& 0.0	& 0.385552401	& -0.0004\\ \hline
O	& 0.0	& 0.5	& 0.38362741	& 0.0	& 0.5	& 0.385552401	& -0.0004\\ \hline
O	& 0.5	& 0.0	& 0.31560125	& 0.5	& 0.0	& 0.318071611	& -0.0004\\ \hline
O	& 0.0	& 0.5	& 0.31560125	& 0.0	& 0.5	& 0.318071611	& -0.0004\\ \hline
O	& 0.5	& 0.0	& 0.24672202	& 0.5	& 0.0	& 0.250461648	& -0.0003\\ \hline
O	& 0.0	& 0.5	& 0.24672202	& 0.0	& 0.5	& 0.250461648	& -0.0003\\ \hline
O	& 0.5	& 0.0	& 0.17632499	& 0.5	& 0.0	& 0.177736344	& 0.0000\\ \hline
O	& 0.0	& 0.5	& 0.17632499	& 0.0	& 0.5	& 0.177736344	& 0.0000\\ \hline
O	& 0.5	& 0.0	& 0.10592795	& 0.5	& 0.0	& 0.105064955	& 0.0017\\ \hline
O	& 0.0	& 0.5	& 0.10592795	& 0.0	& 0.5	& 0.105064955	& 0.0017\\ \hline
O	& 0.5	& 0.0	& 0.0355309	    & 0.5	& 0.0	& 0.031781652	& 0.0000\\ \hline
O	& 0.0	& 0.5	& 0.0355309	    & 0.0	& 0.5	& 0.031781652	& 0.0000\\ \hline
O	& 0.5	& 0.5	& 0.50856393	& 0.5	& 0.5	& 0.284469898	& 0.0023\\ \hline
O	& 0.5	& 0.5	& 0.60685115	& 0.5	& 0.5	& 0.351866946	& -0.0001\\ \hline
O	& 0.5	& 0.5	& 0.70513835	& 0.5	& 0.5	& 0.419225333	& -0.0010\\ \hline
O	& 0.5	& 0.5	& 0.80342556	& 0.5	& 0.5	& 0.486565372	& -0.0012\\ \hline
O	& 0.5	& 0.5	& 0.90171276	& 0.5	& 0.5	& 0.553932208	& -0.0013\\ \hline
O	& 0.5	& 0.5	& 0.00032888	& 0.5	& 0.5	& 0.621412939	& -0.0010\\ \hline
O	& 0.5	& 0.5	& 0.10204167	& 0.5	& 0.5	& 0.688953872	& 0.0006\\ \hline
O	& 0.5	& 0.5	& 0.20375445	& 0.5	& 0.5	& 0.756691147	& 0.0077\\ \hline
N	& 0.5	& 0.5	& 0.30546724	& 0.5	& 0.5	& -0.004579815	& -0.0011\\ \hline
N	& 0.5	& 0.5	& 0.40718003	& 0.5	& 0.5	& 0.068638306	& -0.0006\\ \hline
N	& 0.5	& 0.5	& 0.55770754	& 0.5	& 0.5	& 0.141344443	& 0.0000\\ \hline
N	& 0.5	& 0.5	& 0.65599474	& 0.5	& 0.5	& 0.213907402	& -0.0007\\ \hline
Ti	& 0.5	& 0.5	& 0.75428195	& 0.5	& 0.5	& 0.317383607	& -0.0046\\ \hline
Ti	& 0.5	& 0.5	& 0.85256917	& 0.5	& 0.5	& 0.385021675	& -0.0017\\ \hline
Ti	& 0.5	& 0.5	& 0.95085636	& 0.5	& 0.5	& 0.452423716	& -0.0009\\ \hline
Ti	& 0.5	& 0.5	& 0.51967968	& 0.5	& 0.5	& 0.519760761	& -0.0008\\ \hline
Ti	& 0.5	& 0.5	& 0.58770583	& 0.5	& 0.5	& 0.587149414	& -0.0011\\ \hline
Ti	& 0.5	& 0.5	& 0.65573197	& 0.5	& 0.5	& 0.654599868	& -0.0026\\ \hline
Ti	& 0.5	& 0.5	& 0.7237581	    & 0.5	& 0.5	& 0.721821759	& -0.0101\\ \hline
Ti	& 0.5	& 0.5	& 0.79178425	& 0.5	& 0.5	& 0.788115649	&-0.0078\\ \hline
\end{tabular}
\end{adjustbox}
\caption{Atomic positions and polarization contributions per atom in the V-4x8 Type SrO SC. }
\end{center}
\end{table}
}

{
\begin{table}
\begin{center}
\begin{adjustbox}{width=0.67\textwidth}
\begin{tabular}{|c|c|c|c|c|c|c|}
\hline
\multicolumn{7}{|c|}{V-4x8 Type SrN}\\ \hline						
\multirow{4}{1em}{\bf{A}}	&\multicolumn{3}{|c|}{\bf{CELL PARAMETERS (IN)}}	&	\multicolumn{3}{|c|}{\bf{CELL PARAMETERS (OUT)}} 	\\ \cline{2-7}
& ${a}\, \, (\mathrm{\mbox{{\tiny{\AA}}}} )$ & ${b}\, \,(\mathrm{\mbox{{\tiny{\AA}}}} )$ & ${c}\, \, (\mathrm{\mbox{{\tiny{\AA}}}} )$ & ${a}\, \,(\mathrm{\mbox{{\tiny{\AA}}}} )$ & ${b}\, \, (\mathrm{\mbox{{\tiny{\AA}}}} )$ & ${c}\, \,(\mathrm{\mbox{{\tiny{\AA}}}} )$   \\ \cline{2-7}
	&	3.969	& 3.969 & 	57.404396876 & 	3.9579747444	& 3.9579747444 & 58.270753334	  \\ \cline{2-7}
&	\multicolumn{3}{|c|}{\bf{ATOMIC POSITIONS (crystal)}} & \multicolumn{3}{|c|}{\bf{ATOMIC POSITIONS (crystal)}}		\\ \hline
Sr	& 0.0	& 0.0	& 0.28158819	& 0.0	& 0.0	& 0.286796419\\ \hline
Sr	& 0.0	& 0.0	& 0.34961433	& 0.0	& 0.0	& 0.352465736\\ \hline
Sr	& 0.0	& 0.0	& 0.41764047	& 0.0	& 0.0	& 0.419346471\\ \hline
Sr	& 0.0	& 0.0	& 0.48566661	& 0.0	& 0.0	& 0.48655938\\ \hline
Sr	& 0.0	& 0.0	& 0.55369276	& 0.0	& 0.0	& 0.55388412\\ \hline
Sr	& 0.0	& 0.0	& 0.62171889	& 0.0	& 0.0	& 0.621387879\\ \hline
Sr	& 0.0	& 0.0	& 0.68974504	& 0.0	& 0.0	& 0.689358758\\ \hline
Sr	& 0.0	& 0.0	& 0.75777117	& 0.0	& 0.0	& 0.759074913\\ \hline
Sr	& 0.0	& 0.0	& 0.00000000	& 0.0	& 0.0	& -0.001976023\\ \hline
Sr	& 0.0	& 0.0	& 0.07039704	& 0.0	& 0.0	& 0.068945066\\ \hline
Sr	& 0.0	& 0.0	& 0.14079408	& 0.0	& 0.0	& 0.140926021\\ \hline
Sr	& 0.0	& 0.0	& 0.21119112	& 0.0	& 0.0	& 0.213174489\\ \hline
Ta	& 0.5	& 0.5	& 0.03179161	& 0.5	& 0.5	& 0.033117246\\ \hline
Ta	& 0.5	& 0.5	& 0.10218865	& 0.5	& 0.5	& 0.105571775\\ \hline
Ta	& 0.5	& 0.5	& 0.17258569	& 0.5	& 0.5	& 0.177727693\\ \hline
Ta	& 0.5	& 0.5	& 0.24298273	& 0.5	& 0.5	& 0.24949255\\ \hline
O	& 0.5	& 0.0	& 0.79178425	& 0.5	& 0.0	& 0.789784962\\ \hline
O	& 0.0	& 0.5	& 0.79178425	& 0.0	& 0.5	& 0.789784962\\ \hline
O	& 0.5	& 0.0	& 0.7237581	    & 0.5	& 0.0	& 0.72248547\\ \hline
O	& 0.0	& 0.5	& 0.7237581	    & 0.0	& 0.5	& 0.72248547\\ \hline
O	& 0.5	& 0.0	& 0.65573197	& 0.5	& 0.0	& 0.655092777\\ \hline
O	& 0.0	& 0.5	& 0.65573197	& 0.0	& 0.5	& 0.655092777\\ \hline
O	& 0.5	& 0.0	& 0.58770583	& 0.5	& 0.0	& 0.587663728\\ \hline
O	& 0.0	& 0.5	& 0.58770583	& 0.0	& 0.5	& 0.587663728\\ \hline
O	& 0.5	& 0.0	& 0.51967968	& 0.5	& 0.0	& 0.520304386\\ \hline
O	& 0.0	& 0.5	& 0.51967968	& 0.0	& 0.5	& 0.520304386\\ \hline
O	& 0.5	& 0.0	& 0.45165354	& 0.5	& 0.0	& 0.452980554\\ \hline
O	& 0.0	& 0.5	& 0.45165354	& 0.0	& 0.5	& 0.452980554\\ \hline
O	& 0.5	& 0.0	& 0.38362741	& 0.5	& 0.0	& 0.385607917\\ \hline
O	& 0.0	& 0.5	& 0.38362741	& 0.0	& 0.5	& 0.385607917\\ \hline
O	& 0.5	& 0.0	& 0.31560125	& 0.5	& 0.0	& 0.318090249\\ \hline
O	& 0.0	& 0.5	& 0.31560125	& 0.0	& 0.5	& 0.318090249\\ \hline
O	& 0.5	& 0.0	& 0.24672202	& 0.5	& 0.0	& 0.246234728\\ \hline
O	& 0.0	& 0.5	& 0.24672202	& 0.0	& 0.5	& 0.246234728\\ \hline
O	& 0.5	& 0.0	& 0.17632499	& 0.5	& 0.0	& 0.174485167\\ \hline
O	& 0.0	& 0.5	& 0.17632499	& 0.0	& 0.5	& 0.174485167\\ \hline
O	& 0.5	& 0.0	& 0.10592795	& 0.5	& 0.0	& 0.102335185\\ \hline
O	& 0.0	& 0.5	& 0.10592795	& 0.0	& 0.5	& 0.102335185\\ \hline
O	& 0.5	& 0.0	& 0.0355309	    & 0.5	& 0.0	& 0.030008839\\ \hline
O	& 0.0	& 0.5	& 0.0355309	    & 0.0	& 0.5	& 0.030008839\\ \hline
O	& 0.5	& 0.5	& 0.28158819	& 0.5	& 0.5	& 0.28253055\\ \hline
O	& 0.5	& 0.5	& 0.34961433	& 0.5	& 0.5	& 0.352015025\\ \hline
O	& 0.5	& 0.5	& 0.41764047	& 0.5	& 0.5	& 0.41934174\\ \hline
O	& 0.5	& 0.5	& 0.48566661	& 0.5	& 0.5	& 0.486655276\\ \hline
O	& 0.5	& 0.5	& 0.55369276	& 0.5	& 0.5	& 0.553989698\\ \hline
O	& 0 .5	& 0.5	& 0.62171889	& 0.5	& 0.5	& 0.6214101\\ \hline
O	& 0.5	& 0.5	& 0.68974504	& 0.5	& 0.5	& 0.688906741\\ \hline
O	& 0.5	& 0.5	& 0.75777117	& 0.5	& 0.5	& 0.756549792\\ \hline
N	& 0.5	& 0.5	& 0.00022762	& 0.5	& 0.5	& -0.003626669\\ \hline
N	& 0.5	& 0.5	& 0.07062466	& 0.5	& 0.5	& 0.066225778\\ \hline
N	& 0.5	& 0.5	& 0.14102171	& 0.5	& 0.5	& 0.138502267\\ \hline
N	& 0.5	& 0.5	& 0.21141874	& 0.5	& 0.5	& 0.210625375\\ \hline
Ti	& 0.5	& 0.5	& 0.31560125	& 0.5	& 0.5	& 0.317595841\\ \hline
Ti	& 0.5	& 0.5	& 0.38362741	& 0.5	& 0.5	& 0.385428075\\ \hline
Ti	& 0.5	& 0.5	& 0.45165354	& 0.5	& 0.5	& 0.452856608\\ \hline
Ti	& 0.5	& 0.5	& 0.51967968	& 0.5	& 0.5	& 0.520190794\\ \hline
Ti	& 0.5	& 0.5	& 0.58770583	& 0.5	& 0.5	& 0.587545776\\ \hline
Ti	& 0.5	& 0.5	& 0.65573197	& 0.5	& 0.5	& 0.654907674\\ \hline
Ti	& 0.5	& 0.5	& 0.7237581	    & 0.5	& 0.5	& 0.722010713\\ \hline
Ti	& 0.5	& 0.5	& 0.79178425	& 0.5	& 0.5	& 0.788074812\\ \hline
\end{tabular}
\end{adjustbox}
\caption{Atomic positions in the V-4x8 Type SrN SC. }
\end{center}
\end{table}
}

{
\begin{table}
\begin{center}
\begin{adjustbox}{width=0.67\textwidth}
\begin{tabular}{|c|c|c|c|c|c|c|c|}
\hline
\multicolumn{8}{|c|}{V-5x10 Type SrO}\\ \hline						
\multirow{4}{1em}{\bf{A}}	&\multicolumn{3}{|c|}{\bf{CELL PARAMETERS (IN)}}	&	\multicolumn{3}{|c|}{\bf{CELL PARAMETERS (OUT)}} &	 \multirow{3}{8em}{ {\bf{Born's Effective Charge Method}}} 	\\ \cline{2-7}
& ${a}\, \, (\mathrm{\mbox{{\tiny{\AA}}}} )$ & ${b}\, \,(\mathrm{\mbox{{\tiny{\AA}}}} )$ & ${c}\, \, (\mathrm{\mbox{{\tiny{\AA}}}} )$ & ${a}\, \,(\mathrm{\mbox{{\tiny{\AA}}}} )$ & ${b}\, \, (\mathrm{\mbox{{\tiny{\AA}}}} )$ & ${c}\, \,(\mathrm{\mbox{{\tiny{\AA}}}} )$ &  \\ \cline{2-7}
	&	3.969	& 3.969 & 	69.255495893 & 	3.9557999484 	& 3.9557999484 & 70.3555836411	 & \\ \cline{2-7}
&	\multicolumn{3}{|c|}{\bf{ATOMIC POSITIONS (crystal)}} & \multicolumn{3}{|c|}{\bf{ATOMIC POSITIONS (crystal)}}	&	${(C/m^2)} $	\\ \hline
Sr	& 0.0	& 0.0	& 0.29175302	& 0.0	& 0.0	& 0.295780139	& 0.0000\\ \hline
Sr	& 0.0	& 0.0	& 0.34813843	& 0.0	& 0.0	& 0.350706958	& 0.0000\\ \hline
Sr	& 0.0	& 0.0	& 0.40452385	& 0.0	& 0.0	& 0.406260688	& 0.0000\\ \hline
Sr	& 0.0	& 0.0	& 0.46090925	& 0.0	& 0.0	& 0.461965213	& 0.0000\\ \hline
Sr	& 0.0	& 0.0	& 0.51729466	& 0.0	& 0.0	& 0.517671549	& 0.0000\\ \hline
Sr	& 0.0	& 0.0	& 0.57368008	& 0.0	& 0.0	& 0.573456327	& 0.0000\\ \hline
Sr	& 0.0	& 0.0	& 0.63006549	& 0.0	& 0.0	& 0.629376737	& 0.0000\\ \hline
Sr	& 0.0	& 0.0	& 0.68645089	& 0.0	& 0.0	& 0.68543746	& 0.0000\\ \hline
Sr	& 0.0	& 0.0	& 0.74283631	& 0.0	& 0.0	& 0.741776421	& 0.0000\\ \hline
Sr	& 0.0	& 0.0	& 0.79922172	& 0.0	& 0.0	& 0.79945229	& 0.0000\\ \hline
Sr	& 0.0	& 0.0	& 0.00000000	& 0.0	& 0.0	& -0.005689487	& 0.0000\\ \hline
Sr	& 0.0	& 0.0	& 0.0583506	    & 0.0	& 0.0	& 0.053198144	& 0.0000\\ \hline
Sr	& 0.0	& 0.0	& 0.11670119	& 0.0	& 0.0	& 0.113449696	& 0.0000\\ \hline
Sr	& 0.0	& 0.0	& 0.17505179	& 0.0	& 0.0	& 0.17372083	& 0.0000\\ \hline
Sr	& 0.0	& 0.0	& 0.23340239	& 0.0	& 0.0	& 0.234098962	& 0.0000\\ \hline
Ta	& 0 .5	& 0.5	& 0.02635138	& 0.5	& 0.5	& 0.022003935	& -0.0005\\ \hline
Ta	& 0.5	& 0.5	& 0.08470198	& 0.5	& 0.5	& 0.082789326	& 0.0001\\ \hline
Ta	& 0.5	& 0.5	& 0.14305258	& 0.5	& 0.5	& 0.143107301	& 0.0001\\ \hline
Ta	& 0.5	& 0.5	& 0.20140318	& 0.5	& 0.5	& 0.203321562	& 0.0000\\ \hline
Ta	& 0.5	& 0.5	& 0.25975378	& 0.5	& 0.5	& 0.263369916	& -0.0022\\ \hline
O	& 0.5	& 0.0	& 0.31994573	& 0.5	& 0.0	& 0.322843477	& 0.0004\\ \hline
O	& 0.0	& 0.5	& 0.31994573	& 0.0	& 0.5	& 0.322843477	& 0.0004\\ \hline
O	& 0.5	& 0.0	& 0.37633114	& 0.5	& 0.0	& 0.37870477	& -0.0003\\ \hline
O	& 0.0	& 0.5	& 0.37633114	& 0.0	& 0.5	& 0.37870477	& -0.0003\\ \hline
O	& 0.5	& 0.0 	& 0.43271655	& 0.5	& 0.0	& 0.434495988	& -0.0005\\ \hline
O	& 0.0	& 0.5	& 0.43271655	& 0.0	& 0.5	& 0.434495988	& -0.0005\\ \hline
O	& 0.5	& 0.0	& 0.48910196	& 0.5	& 0.0	& 0.490198193	& -0.0005\\ \hline
O	& 0.0	& 0.5	& 0.48910196	& 0.0	& 0.5	& 0.490198193	& -0.0006\\ \hline
O	& 0.5	& 0.0	& 0.54548737	& 0.5	& 0.0	& 0.545921293	& -0.0005\\ \hline
O	& 0.0	& 0.5	& 0.54548737	& 0.0	& 0.5	& 0.545921293	& -0.0005\\ \hline
O	& 0.5	& 0.0	& 0.60187278	& 0.5	& 0.0	& 0.601753108	& -0.0005\\ \hline
O	& 0.0	& 0.5	& 0.60187278	& 0.0	& 0.5	& 0.601753108	& -0.0005\\ \hline
O	& 0.5	& 0.0	& 0.6582582	    & 0.5	& 0.0	& 0.657708597	& -0.0005\\ \hline
O	& 0.0	& 0.5	& 0.6582582	    & 0.0	& 0.5	& 0.657708597	& -0.0005\\ \hline
O	& 0.5	& 0.0	& 0.7146436	    & 0.5	& 0.0	& 0.713760649	& -0.0002\\ \hline
O	& 0.0	& 0.5	& 0.7146436	    & 0.0	& 0.5	& 0.713760649	& -0.0002\\ \hline
O	& 0.5	& 0.0	& 0.77102901	& 0.5	& 0.0	& 0.769764438	& 0.0013\\ \hline
O	& 0.0	& 0.5	& 0.77102901	& 0.0	& 0.5	& 0.769764438	& 0.0013\\ \hline
O	& 0.5	& 0.0	& 0.82741443	& 0.5	& 0.0	& 0.825600931	& 0.0000\\ \hline
O	& 0.0	& 0.5	& 0.82741443	& 0.0	& 0.5	& 0.825600931	& 0.0000\\ \hline
O	& 0.5	& 0.0	& 0.0294508	    & 0.5	& 0.0	& 0.025141821	& -0.0002\\ \hline
O	& 0.0	& 0.5	& 0.0294508	    & 0.0	& 0.5	& 0.025141821	& -0.0002\\ \hline
O	& 0.5	& 0.0	& 0.0878014	    & 0.5	& 0.0	& 0.086047707	& 0.0000\\ \hline
O	& 0.0	& 0.5	& 0.0878014	    & 0.0	& 0.5	& 0.086047707	& 0.0000\\ \hline
O	& 0.5	& 0.0	& 0.146152		& 0.5	& 0.0	& 0.14636021	& 0.0000\\ \hline
O	& 0.0	& 0.5	& 0.146152		& 0.0	& 0.5	& 0.14636021	& 0.0000\\ \hline
O	& 0.5	& 0.0	& 0.2045026	    & 0.5	& 0.0	& 0.206580911	& -0.0001\\ \hline
O	& 0.0	& 0.5	& 0.2045026	    & 0.0	& 0.5	& 0.206580911	& -0.0001\\ \hline
O	& 0.5	& 0.0	& 0.2628532	    & 0.5	& 0.0	& 0.266888221	& -0.0008\\ \hline
O	& 0.0	& 0.5	& 0.2628532	    & 0.0	& 0.5	& 0.266888221	& -0.0008\\ \hline
O	& 0.5	& 0.5	& 0.29175302	& 0.5	& 0.5	& 0.295030379	& 0.0017\\ \hline
O	& 0.5	& 0.5	& 0.34813843	& 0.5	& 0.5	& 0.350855463	& -0.0005\\ \hline
O	& 0.5	& 0.5	& 0.40452385	& 0.5	& 0.5	& 0.406615352	& -0.0014\\ \hline
O	& 0.5	& 0.5	& 0.46090925	& 0.5	& 0.5	& 0.462352263	& -0.0016\\ \hline
O	& 0.5	& 0.5	& 0.51729466	& 0.5	& 0.5	& 0.518074507	& -0.0017\\ \hline
O	& 0.5	& 0.5	& 0.57368008	& 0.5	& 0.5	& 0.573860934	& -0.0018\\ \hline
O	& 0.5	& 0.5	& 0.63006549	& 0.5	& 0.5	& 0.629769657	& -0.0017\\ \hline
O	& 0.5	& 0.5	& 0.68645089	& 0.5	& 0.5	& 0.685768332	& -0.0014\\ \hline
O	& 0.5	& 0.5	& 0.74283631	& 0.5	& 0.5	& 0.741796681	& -0.0001\\ \hline
O	& 0.5	& 0.5	& 0.79922172	& 0.5	& 0.5	& 0.797934295	& 0.0063\\ \hline
N	& 0.5	& 0.5	& 0.00018867	& 0.5	& 0.5	& -0.004940197	& -0.0010\\ \hline
N	& 0.5	& 0.5	& 0.05853927	& 0.5	& 0.5	& 0.055814102	& -0.0006\\ \hline
N	& 0.5	& 0.5	& 0.11688986	& 0.5	& 0.5	& 0.116144657	& 0.0003\\ \hline
N	& 0.5	& 0.5	& 0.17524046	& 0.5	& 0.5	& 0.17637923	& 0.0002\\ \hline
N	& 0.5	& 0.5	& 0.23359106	& 0.5	& 0.5	& 0.236558402	& -0.0006\\ \hline
Ti	& 0.5	& 0.5	& 0.31994573	& 0.5	& 0.5	& 0.322175866	& -0.0041\\ \hline
Ti	& 0.5	& 0.5	& 0.37633114	& 0.5	& 0.5	& 0.378185814	& -0.0014\\ \hline
Ti	& 0.5	& 0.5	& 0.43271655	& 0.5	& 0.5	& 0.433990622	& -0.0006\\ \hline
Ti	& 0.5	& 0.5	& 0.48910196	& 0.5	& 0.5	& 0.489714063	& -0.0006\\ \hline
Ti	& 0.5	& 0.5	& 0.54548737	& 0.5	& 0.5	& 0.545446815	& -0.0006\\ \hline
Ti	& 0.5	& 0.5	& 0.60187278	& 0.5	& 0.5	& 0.601273759	& -0.0008\\ \hline
Ti	& 0.5	& 0.5	& 0.6582582	    & 0.5	& 0.5	& 0.657217301	& -0.0010\\ \hline
Ti	& 0.5	& 0.5	& 0.7146436	    & 0.5	& 0.5	& 0.713194228	& -0.0023\\ \hline
Ti	& 0.5	& 0.5	& 0.77102901	& 0.5	& 0.5	& 0.768961222	& -0.0088\\ \hline
Ti	& 0.5	& 0.5	& 0.82741443	& 0.5	& 0.5	& 0.823919556	&-0.0074\\ \hline
\end{tabular}
\end{adjustbox}
\caption{Atomic positions and polarization contributions per atom in the V-5x10 Type SrO SC. }
\end{center}
\end{table}
}

{
\begin{table}
\begin{center}
\begin{adjustbox}{width=0.54 \textwidth}
\begin{tabular}{|c|c|c|c|c|c|c|}
\hline
\multicolumn{7}{|c|}{V-5x10 Type SrN}\\ \hline						
\multirow{4}{1em}{\bf{A}}	&\multicolumn{3}{|c|}{\bf{CELL PARAMETERS (IN)}}	&	\multicolumn{3}{|c|}{\bf{CELL PARAMETERS (OUT)}} 	\\ \cline{2-7}
& ${a}\, \, (\mathrm{\mbox{{\tiny{\AA}}}} )$ & ${b}\, \,(\mathrm{\mbox{{\tiny{\AA}}}} )$ & ${c}\, \, (\mathrm{\mbox{{\tiny{\AA}}}} )$ & ${a}\, \,(\mathrm{\mbox{{\tiny{\AA}}}} )$ & ${b}\, \, (\mathrm{\mbox{{\tiny{\AA}}}} )$ & ${c}\, \,(\mathrm{\mbox{{\tiny{\AA}}}} )$   \\ \cline{2-7}
	&	3.969	& 3.969 & 	70.353 & 	3.959435465 & 3.959435465 & 70.337002752  \\ \cline{2-7}
&	\multicolumn{3}{|c|}{\bf{ATOMIC POSITIONS (crystal)}} & \multicolumn{3}{|c|}{\bf{ATOMIC POSITIONS (crystal)}}		\\ \hline
Sr	& 0.0	& 0.0	& 0.29175302	& 0.0	& 0.0	& 0.296967173\\ \hline
Sr	& 0.0	& 0.0	& 0.34813843	& 0.0	& 0.0	& 0.350977768\\ \hline
Sr	& 0.0	& 0.0	& 0.40452385	& 0.0	& 0.0	& 0.406038902\\ \hline
Sr	& 0.0	& 0.0	& 0.46090925	& 0.0	& 0.0	& 0.46152718\\ \hline
Sr	& 0.0	& 0.0	& 0.51729466	& 0.0	& 0.0	& 0.517508933\\ \hline
Sr	& 0.0	& 0.0	& 0.57368008	& 0.0	& 0.0	& 0.573752579\\ \hline
Sr	& 0.0	& 0.0	& 0.63006549	& 0.0	& 0.0	& 0.63006129\\ \hline
Sr	& 0.0	& 0.0	& 0.68645089	& 0.0	& 0.0	& 0.686323663\\ \hline
Sr	& 0.0	& 0.0	& 0.74283631	& 0.0	& 0.0	& 0.742580965\\ \hline
Sr	& 0.0	& 0.0	& 0.79922172	& 0.0	& 0.0	& 0.799951067\\ \hline
Sr	& 0.0	& 0.0	& 0.00000000	& 0.0	& 0.0	& -0.001909218\\ \hline
Sr	& 0.0	& 0.0	& 0.0583506	    & 0.0	& 0.0	& 0.056748129\\ \hline
Sr	& 0.0	& 0.0	& 0.11670119	& 0.0	& 0.0	& 0.116311362\\ \hline
Sr	& 0.0	& 0.0	& 0.17505179	& 0.0	& 0.0	& 0.176032751\\ \hline
Sr	& 0.0	& 0.0	& 0.23340239	& 0.0	& 0.0	& 0.236015946\\ \hline
Ta	& 0.5	& 0.5	& 0.02635138	& 0.5	& 0.5	& 0.026950245\\ \hline
Ta	& 0.5	& 0.5	& 0.08470198	& 0.5	& 0.5	& 0.086955665\\ \hline
Ta	& 0.5	& 0.5	& 0.14305258	& 0.5	& 0.5	& 0.146741382\\ \hline
Ta	& 0.5	& 0.5	& 0.20140318	& 0.5	& 0.5	& 0.206459912\\ \hline
Ta	& 0.5	& 0.5	& 0.25975378	& 0.5	& 0.5	& 0.265982287\\ \hline
O	& 0.5	& 0.0	& 0.31994573	& 0.5	& 0.0	& 0.32271966\\ \hline
O	& 0.0	& 0.5	& 0.31994573	& 0.0	& 0.5	& 0.32271966\\ \hline
O	& 0.5	& 0.0	& 0.37633114	& 0.5	& 0.0	& 0.378202029\\ \hline
O	& 0.0	& 0.5	& 0.37633114	& 0.0	& 0.5	& 0.378202029\\ \hline
O	& 0.5	& 0.0	& 0.43271655	& 0.5	& 0.0	& 0.43358976\\ \hline
O	& 0.0	& 0.5	& 0.43271655	& 0.0	& 0.5	& 0.43358976\\ \hline
O	& 0.5	& 0.0	& 0.48910196	& 0.5	& 0.0	& 0.489357379\\ \hline
O	& 0.0	& 0.5	& 0.48910196	& 0.0	& 0.5	& 0.489357379\\ \hline
O	& 0.5	& 0.0	& 0.54548737	& 0.5	& 0.0	& 0.545529378\\ \hline
O	& 0.0	& 0.5	& 0.54548737	& 0.0	& 0.5	& 0.545529378\\ \hline
O	& 0.5	& 0.0	& 0.60187278	& 0.5	& 0.0	& 0.601841163\\ \hline
O	& 0.0	& 0.5	& 0.60187278	& 0.0	& 0.5	& 0.601841163\\ \hline
O	& 0.5	& 0.0	& 0.6582582	    & 0.5	& 0.0	& 0.658102109\\ \hline
O	& 0.0	& 0.5	& 0.6582582	    & 0.0	& 0.5	& 0.658102109\\ \hline
O	& 0.5	& 0.0	& 0.7146436	    & 0.5	& 0.0	& 0.714105889\\ \hline
O	& 0.0	& 0.5	& 0.7146436	    & 0.0	& 0.5	& 0.714105889\\ \hline
O	& 0.5	& 0.0	& 0.77102901	& 0.5	& 0.0	& 0.769809612\\ \hline
O	& 0.0	& 0.5	& 0.77102901	& 0.0	& 0.5	& 0.769809612\\ \hline
O	& 0.5	& 0.0	& 0.82741443	& 0.5	& 0.0	& 0.825530281\\ \hline
O	& 0.0	& 0.5	& 0.82741443	& 0.0	& 0.5	& 0.825530281\\ \hline
O	& 0.5	& 0.0	& 0.0294508	    & 0.5	& 0.0	& 0.02441445\\ \hline
O	& 0.0	& 0.5	& 0.0294508	    & 0.0	& 0.5	& 0.02441445\\ \hline
O	& 0.5	& 0.0	& 0.0878014	    & 0.5	& 0.0	& 0.084142417\\ \hline
O	& 0.0	& 0.5	& 0.0878014	    & 0.0	& 0.5	& 0.084142417\\ \hline
O	& 0.5	& 0.0	& 0.146152		& 0.5	& 0.0	& 0.143826505\\ \hline
O	& 0.0	& 0.5	& 0.146152		& 0.0	& 0.5	& 0.143826505\\ \hline
O	& 0.5	& 0.0	& 0.2045026	    & 0.5	& 0.0	& 0.203571124\\ \hline
O	& 0.0	& 0.5	& 0.2045026	    & 0.0	& 0.5	& 0.203571124\\ \hline
O	& 0.5	& 0.0	& 0.2628532	    & 0.5	& 0.0	& 0.263166258\\ \hline
O	& 0.0	& 0.5	& 0.2628532	    & 0.0	& 0.5	& 0.263166258\\ \hline
O	& 0.5	& 0.5	& 0.34813843	& 0.5	& 0.5	& 0.350457238\\ \hline
O	& 0.5	& 0.5	& 0.40452385	& 0.5	& 0.5	& 0.405914571\\ \hline
O	& 0.5	& 0.5	& 0.46090925	& 0.5	& 0.5	& 0.461539722\\ \hline
O	& 0.5	& 0.5	& 0.51729466	& 0.5	& 0.5	& 0.517494611\\ \hline
O	& 0.5	& 0.5	& 0.57368008	& 0.5	& 0.5	& 0.573710817\\ \hline
O	& 0.5	& 0.5	& 0.63006549	& 0.5	& 0.5	& 0.629990276\\ \hline
O	& 0.5	& 0.5	& 0.68645089	& 0.5	& 0.5	& 0.686128947\\ \hline
O	& 0.5	& 0.5	& 0.74283631	& 0.5	& 0.5	& 0.742023746\\ \hline
O	& 0.5	& 0.5	& 0.79922172	& 0.5	& 0.5	& 0.797959657\\ \hline
O	& 0.5	& 0.5	& 0.00018867	& 0.5	& 0.5	& -0.003372075\\ \hline
N	& 0.5	& 0.5	& 0.05853927	& 0.5	& 0.5	& 0.05430743\\ \hline
N	& 0.5	& 0.5	& 0.11688986	& 0.5	& 0.5	& 0.11413344\\ \hline
N	& 0.5	& 0.5	& 0.17524046	& 0.5	& 0.5	& 0.17388814\\ \hline
N	& 0.5	& 0.5	& 0.23359106	& 0.5	& 0.5	& 0.233615404\\ \hline
N	& 0.5	& 0.5	& 0.29175302	& 0.5	& 0.5	& 0.293220176\\ \hline
Ti	& 0.5	& 0.5	& 0.31994573	& 0.5	& 0.5	& 0.322366487\\ \hline
Ti	& 0.5	& 0.5	& 0.37633114	& 0.5	& 0.5	& 0.378130544\\ \hline
Ti	& 0.5	& 0.5	& 0.43271655	& 0.5	& 0.5	& 0.433644536\\ \hline
Ti	& 0.5	& 0.5	& 0.48910196	& 0.5	& 0.5	& 0.489402621\\ \hline
Ti	& 0.5	& 0.5	& 0.54548737	& 0.5	& 0.5	& 0.545490801\\ \hline
Ti	& 0.5	& 0.5	& 0.60187278	& 0.5	& 0.5	& 0.601744687\\ \hline
Ti	& 0.5	& 0.5	& 0.6582582	    & 0.5	& 0.5	& 0.657947717\\ \hline
Ti	& 0.5	& 0.5	& 0.7146436	    & 0.5	& 0.5	& 0.713861422\\ \hline
Ti	& 0.5	& 0.5	& 0.77102901	& 0.5	& 0.5	& 0.769389892\\ \hline
Ti	& 0.5	& 0.5	& 0.82741443	& 0.5	& 0.5	& 0.824103085\\ \hline
\end{tabular}
\end{adjustbox}
\caption{Atomic positions in the V-5x10 Type SrN SC. }
\end{center}
\end{table}
}
\end{document}